\documentclass{elsart}

\usepackage{epsfig}
\usepackage{amssymb,amsmath}

\graphicspath{{./}}

\begin{document}

\begin{frontmatter}

\title{Optimized network structure \\ 
       and routing metric \\
       in wireless multihop ad hoc communication}

\author[label1,label2]{Wolfram Krause}
  \ead{krause@figss.uni-frankfurt.de}
\author[label3]{Jan Scholz}
  \ead{Jan.C.Scholz@physik.uni-giessen.de}
\author[label1]{Martin Greiner}
  \ead{martin.greiner@siemens.com}
\address[label1]{Corporate Technology, Information \& Communications, 
                 Siemens AG,\\
                 D-81730 M\"unchen, Germany}
\address[label2]{Frankfurt Institute for Advanced Studies and\\
                 Frankfurt International Graduate School for Science,\\
                 Johann Wolfgang Goethe Universit\"at,
                 Max-von-Laue-Straße 1,\\
                 D-60438 Frankfurt am Main, Germany}
\address[label3]{Institut f\"ur Theoretische Physik,\\
                 Justus Liebig Universit\"at,
                 Heinrich-Buff-Ring 16, \\
                 D-35392 Gie{\ss}en, Germany}

\begin{abstract}
Inspired by the Statistical Physics of complex networks, wireless
multihop ad hoc communication networks are considered in abstracted
form. Since such engineered networks are able to modify their structure 
via topology control, we search for optimized network structures, which 
maximize the end-to-end throughput performance. A modified version of 
betweenness centrality is introduced and shown to be very relevant for 
the respective modeling. The calculated optimized network structures 
lead to a significant increase of the end-to-end throughput. The 
discussion of the resulting structural properties reveals that it will 
be almost impossible to construct these optimized topologies in a 
technologically efficient distributive manner. However, the modified 
betweenness centrality also allows to propose a new routing metric for 
the end-to-end communication traffic. This approach leads to an even 
larger increase of throughput capacity and is easily implementable in a
technologically relevant manner. 
\end{abstract}


\end{frontmatter}

\newpage
\section{Introduction}
\label{sec:intro}

Nowadays, the complexity of many engineered networks has increased to
such a high level, that conceptually new approaches for their 
operational control have to be looked for. Key aspects like 
selforganization and artificial intelligence become increasingly 
important. It is here where engineering and computer science is 
beginning to exchange ideas and concepts with the natural sciences like 
physics and biology. In particular, the new cross-disciplinary branch 
known as the Statistical Physics of complex networks 
\cite{alb02,dor03,new03} appears to catalyze such efforts. 

We pick up on this latest momentum to continue the discussion of a 
challenging complex communication network in abstracted form 
\cite{gla03a,kra04,gla04}. In wireless multihop ad hoc networks 
\cite{mob02,mob03,mob04} nodes are connected by wireless links. 
A central control infrastructure is missing. Each node does not only act 
as a communication source and sink, but also forwards communication for 
others. All of this requires a lot of selforganized and decentralized 
coordination amongst the nodes. The outstanding complexity of this 
communication network is revealed by mentioning the key mechanisms and 
the associated problems in some more detail:

With regulation of its transmission power, each node is able to modify 
its transmission range and its neighborhood, to which it builds up
wireless communication links. Here the node faces frustration for the 
first time. On the one hand it wants to save energy and to keep its 
transmission power as low as possible, but on the other it might have to 
choose a larger neighborhood in order to help the network to gain strong 
connectivity, so that each node will be able to communicate to any other 
via multihop routes. This brings us to another protocol layer, from link 
control to routing control. End-to-end routes have to be explored and 
maintained. During their execution the communication hops from one node 
to the next. This is where yet another protocol layer, called medium 
access control, sets in. It blocks all neighbors attached to an ongoing 
one-hop transmission in order to avoid interference within the same 
communication channel. This is the origin of another frustration, now 
across layers. Whereas routing efficiency prefers short end-to-end 
routes with the consequence of large one-hop neighborhoods, medium 
access control prefers to block small neighborhoods with the consequence 
of long end-to-end routes. A delicate balance between these two layers 
is necessary for the overall network to gain a large end-to-end 
throughput capacity, which measures the amount of communication traffic 
the network is able to handle without overloading.

In a previous paper \cite{gla03a} we have already addressed the 
connectivity issue, with special emphasis on the development of a simple 
selforganizing topology control. More selforganization has been proposed 
in Ref.\ \cite{gla04}, where a reactive routing \& congestion control 
has been discussed, which adapts to the current congestion state of the
wireless multihop ad hoc network. Another investigation \cite{kra04} has 
demonstrated that the end-to-end throughput capacity does sensitively 
depend on the underlying network structure. It is exactly here where we 
continue and begin to ask for throughput optimization: What is the 
optimized network structure? What are its properties? Is it possible to 
construct the optimized network structure with a selforganizing topology 
control? Are there also other means to enhance the end-to-end throughput 
and how do these compare to the approach with optimized network 
structures?

These are a lot of questions. We group them into the following 
organizational form of the Paper. Section 2 addresses the network 
structure optimization of wireless multihop ad hoc communication 
networks. It explains the level of abstraction needed to construct an 
objective function for the global maximization of end-to-end throughput. 
A modified version of betweenness centrality is introduced, the  
cumulative betweenness centrality, which suites well the particular 
throughput needs of wireless multihop communication. The optimality of 
the resulting networks is checked with generic packet traffic 
simulations. Scaling laws for the end-to-end throughput with respect to 
network size are given. The structure of the globally optimized networks 
is analyzed and found to be difficult to construct with a decentralized, 
selforganizing topology control. As a consequence, a new approach is 
advocated in Section 3 to increase the end-to-end throughput. It uses 
the cumulative betweenness centrality as key input into a routing 
metric. This allows to determine throughput-optimizing end-to-end routes 
iteratively in a selforganizing manner. The increase in throughput turns 
out to become even slightly larger than for the optimized network 
structure of the previous Section. Conclusion and outlook are given in 
Section 4.

\section{Optimization of network structure}
\label{sec:topology}

\subsection{Abstraction: geometric minimum-node-degree networks}
\label{subsec:twoA}

Some level of abstraction is required to make wireless multihop ad hoc 
communication amenable to the Statistical Physics of complex networks.
The first simplification is to neglect mobility and to distribute $N$
nodes onto a unit square in a random homogeneous way. The transmission 
power $P_i$ then decides which other nodes $j$ are able to be reached by
node $i$ via directed links $i\rightarrow j$. These are the nodes which, 
according to a simple propagation-receiver model, fulfill the inequality 
$P_{i}/R_{ij}^\alpha \geq \mbox{\sc snr}$. $R_{ij}$ denotes the relative 
Euclidean distance. The path-loss exponent $\alpha$ is assumed to be 
constant and $\mbox{\sc snr}$ represents the signal-to-noise ratio. With 
these simplifications, wireless mulithop ad hoc communication networks 
can be modeled as geometric graphs. The $N$ nodes are networked together 
via the set $\{ i\rightarrow j \}$ of all directed links. 

Not all of the directed links will be used for wireless multihop ad hoc
communication. Since operation requires instant one-hop feedback, only 
bidirected links $i \leftrightarrow j$ qualify for the routing of 
communication traffic. It is the bidirected links attached to node $i$ 
which define its communication neighborhood $\mathcal{N}_i$ and its node 
degree $k_i$. 
 
One further step is needed to fully specify wireless multihop ad hoc 
network graphs: assignment of the transmission power $P_i$ for all 
nodes. The simplest procedure is to assign the same value $P$ to all
nodes \cite{gup98,gup00,bet02,dou02}. We prefer to employ a different 
procedure \cite{gla03a}, which contrary to the first one is 
distributive, selforganizing and adaptive. In a nutshell, each node $i$ 
forces its $k_\mathrm{min}$ closest nodes $j$ to adjust their 
transmission powers to at least $P_j=R_{ij}^\alpha$, while adopting the 
value $P_i=\sup_j P_j$ for itself. Its own value can be increased 
further whenever another, still close-by node, which seeks for its 
minimum communication neighborhood, forces $i$ in return to have an even 
larger transmission power. In this respect each node has at least 
$k_\mathrm{min}$ bidirected neighbors. As a consequence transmission 
power values differ from node to node. This heterogeneity leads to the 
occasional emergence of directed links.

\begin{figure}
\begin{center}
\epsfig{file=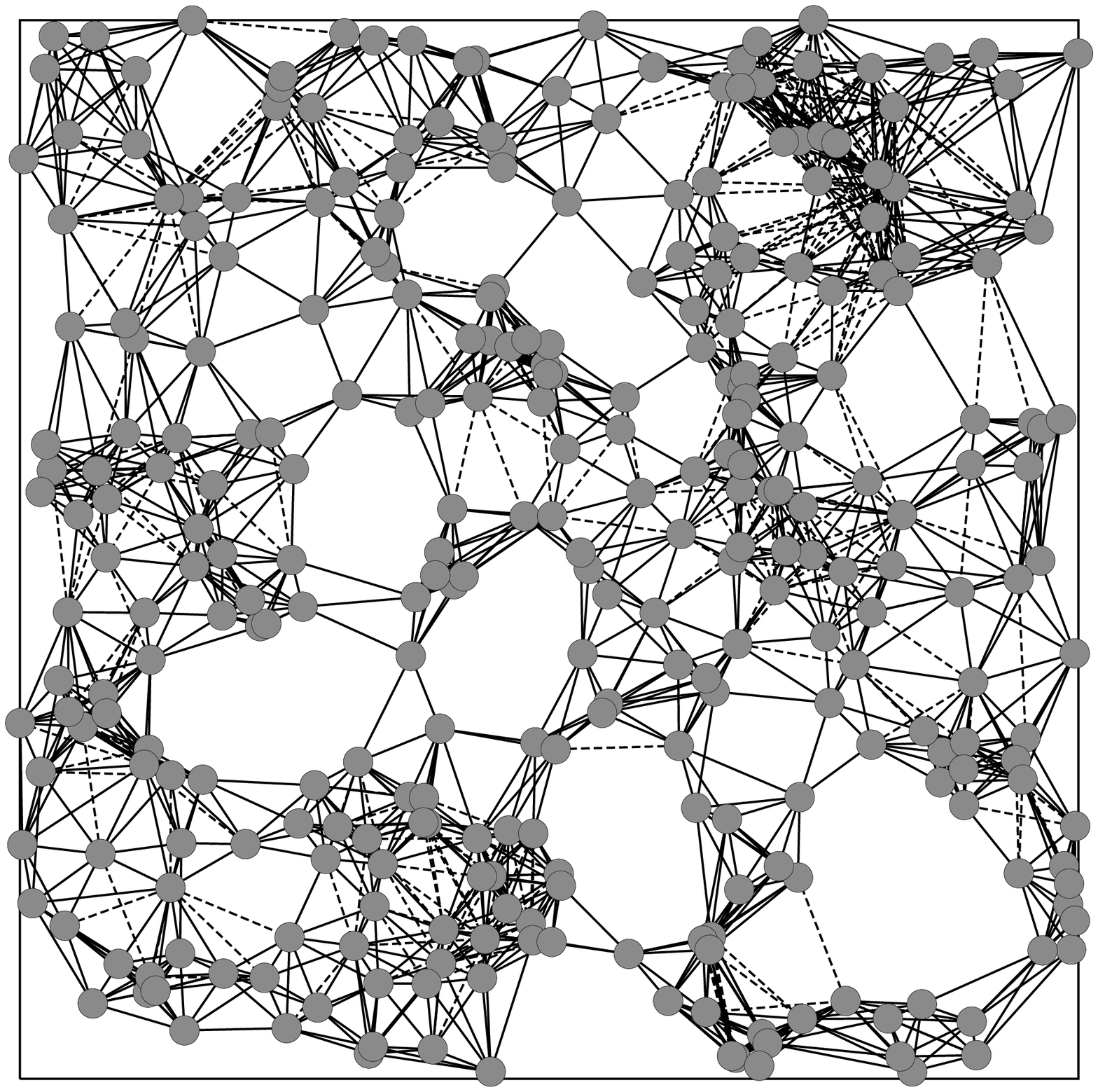,width=0.50\textwidth}
\epsfig{file=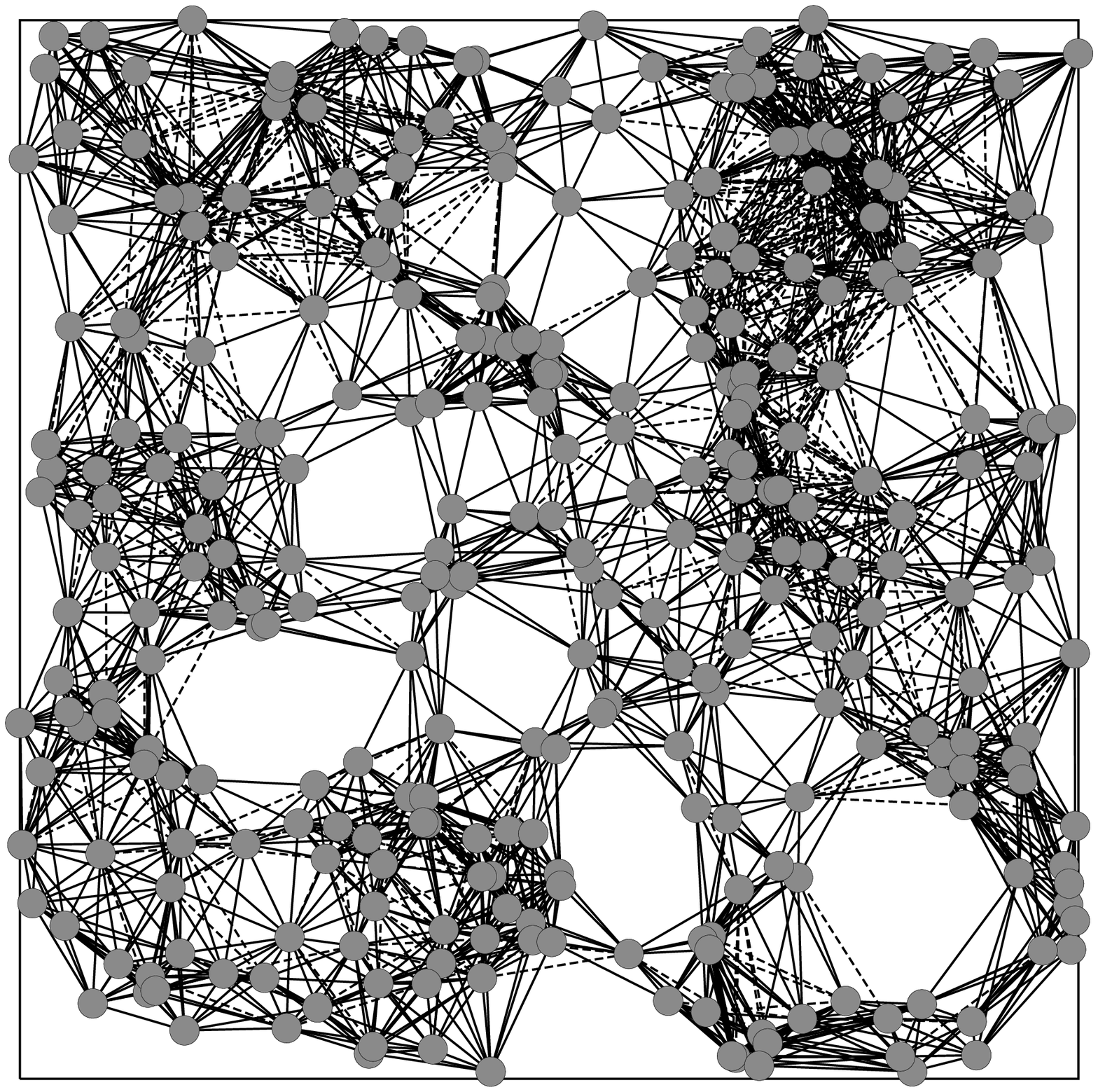,width=0.50\textwidth}
\epsfig{file=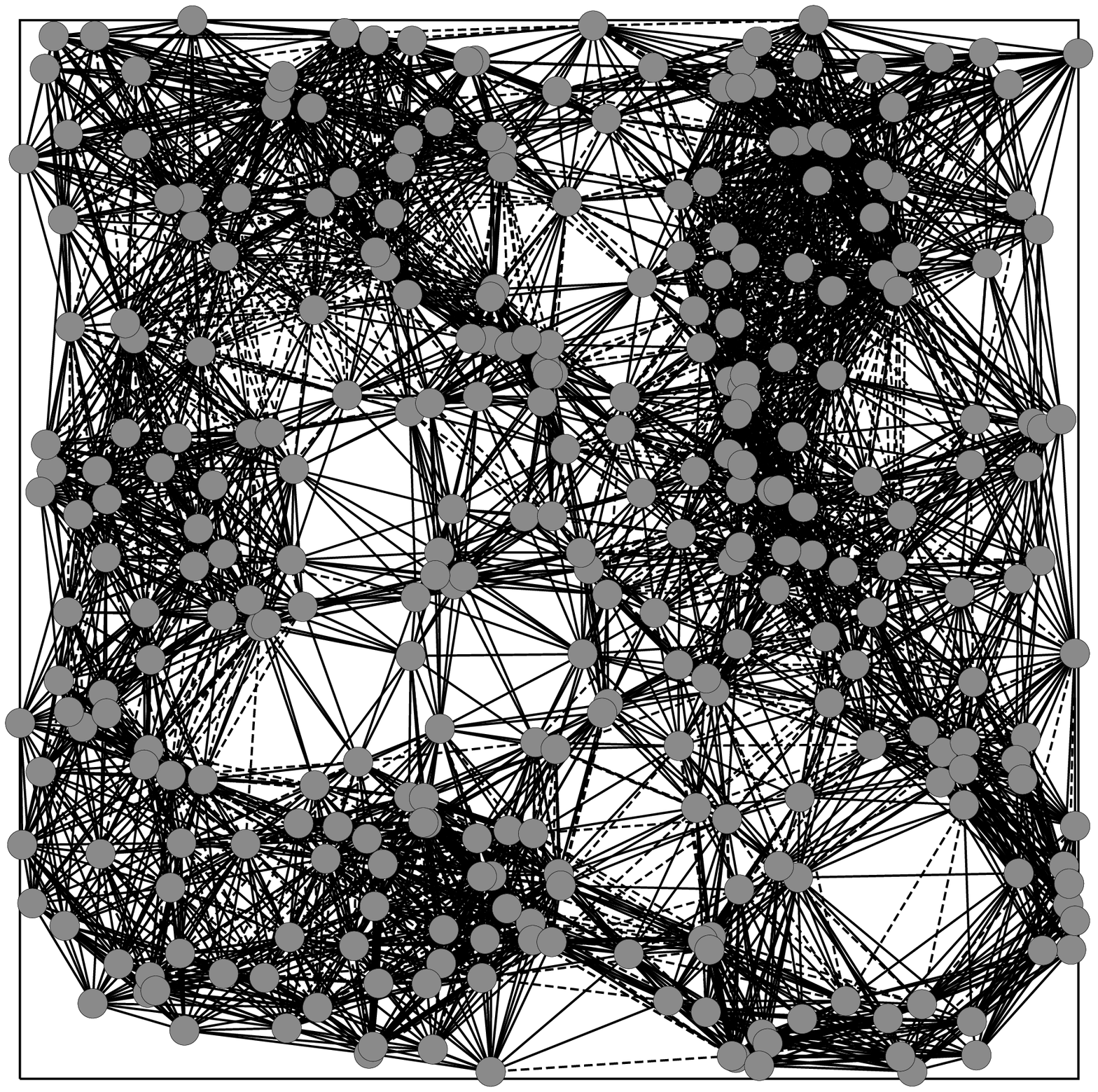,width=0.50\textwidth}
\caption{
Geometric wireless multihop ad hoc network graphs of minimum-node-degree 
type: (a) $k_{\rm min} = 8$, (b) $12$ and (c) $20$. The same random 
homogeneous spatial node pattern with $N=300$ nodes has been used for 
all three cases. Bidirected links are shown in solid, whereas dashed 
links exist only in one direction.
}
\label{fig:fig1}
\end{center}
\end{figure}

We will call wireless multihop ad hoc network graphs generated with this 
heterogeneous power assignment as geometric minimum-node-degree networks. 
Already the choice $k_\mathrm{min}=8$ is sufficient to guarantee strong 
network connectivity almost surely for network sizes up to several 
thousand nodes \cite{gla03a}. A realization of a minimum-node-degree 
network with $k_\mathrm{min}=8$ is shown in Fig.\ \ref{fig:fig1}. For 
comparison, network realizations generated with the same spatial node
pattern, but with $k_\mathrm{min}=12$ and $20$ are also shown.

\subsection{Objective function for end-to-end throughput}
\label{subsec:twoB}

Maybe the most important performance measure of wireless multihop ad
hoc networks is given by the end-to-end throughput capacity $T_{e2e}$. 
It represents the amount of end-to-end communication traffic the network 
is able to handle without overloading. Ref.\ \cite{kra04} has shown that
$T_{e2e}$ does depend on the underlying network structure. We are now 
interested to find network structures with increased throughput capacity.
The necessary first step for this search is the construction of a
suitable objective function in analytical form
that can be calculated with reasonable computational effort.

For simplicity, the end-to-end packet traffic is assumed to use only one
communication channel and to be completely random, that is on average all 
possible $N(N-1)$ end-to-end routes are equally loaded. With rate $\mu$ 
each of the $N$ nodes creates a new packet, for which it selects a random 
final destination, and then sends the packet along the respective 
end-to-end route. As a consequence, the in-packet flux rate 
$\mu_i^{\rm in}$ of end-to-end communications which hop via node $i$ can 
be described as  
\begin{equation}
\label{eq:muiin}
  \mu_i^{\rm in} 
    =  \mu N \frac{B_i}{N(N-1)}
       \; .
\end{equation}
Out of all $N(N-1)$ end-to-end combinations the betweenness centrality
\begin{equation}
\label{eq:bi}
  B_i  =  \mathop{\sum_{m{\neq}n=1}^N}_{(n{\neq}i)}
          \frac{b_{mn}(i)}{b_{mn}}
\end{equation}
counts the number of end-to-end routes going through $i$. To be more 
precise, $B_i$ counts the number of end-to-end routes, for which $i$ has 
to forward a packet. If $i$ is the final recipient of such an end-to-end
route, then $i$ is excluded from the count. Whereas if $i$ is the 
initial sender, then it is included into the count. $b_{mn}$ 
represents the number of used routes from $m$ to $n$, of which 
$b_{mn}(i)$ pass through $i$. 

Betweenness centrality has already been introduced for networks in 
general \cite{new01b} and communication as well as power networks in 
particular \cite{gui02,motter04}. Throughout this section we will base
it on shortest multihop routes. Note however, that betweenness 
centrality is not restricted to the concept of shortest paths 
\cite{new03e}. Other generalizations, i.e.\ from a node-based to a 
link-based betweenness centrality, have also been discussed 
\cite{kra04,motter04}.

Compared to the in-packet flux rate $\mu_i^{\rm in}$, the modeling of
the out-packet flux rate 
\begin{equation}
\label{eq:muiout}
  \mu_i^{\rm out} 
    =  \frac{1}{\tau_i^{\rm send}}
\end{equation}
is more delicate. Given that node $i$ has buffered at least one packet 
to forward, $\tau_i^{\rm send}$ represents the characteristic time it 
takes on average to send the next packet. Due to the competition between 
neighbors to gain medium access for one-hop transmissions, node $i$ 
might not be allowed to send a packet right away. If neighbors also have 
packets to forward or are prospected one-hop receivers, and if one of 
them succeeds, then this node automatically blocks all its neighbors, 
including $i$, by the interference-avoiding medium access control. 

Before we propose an empirical modeling of the sending time 
$\tau_i^{\rm send}$, we write down the general expression for the 
end-to-end throughput. Equating relations (\ref{eq:muiin}) and 
(\ref{eq:muiout}) yields an expression for the critical packet creation 
rate of node $i$,
\begin{equation}
\label{eq:muicrit}
  \mu_i^{\rm crit} 
    =  \frac{N-1}{B_i \tau_i^{\rm send}}
       \; .
\end{equation}
For $\mu < \mu_i^{\rm crit}$ node $i$ is able to forward its incoming 
packet traffic in time and remains subcritical. However, above 
$\mu_i^{\rm crit}$ not all packets can be forwarded, leading to
an increase of its packet queue with time. This defines its 
supercritical regime. Out of all $N$ nodes, it is the node with the
smallest critical packet creation rate, which determines the overall
critical network load
\begin{equation}
\label{eq:mucrit}
  \mu^{\rm crit} 
    =  \mathrm{min}_i \, \mu_i^{\rm crit}
       \; .
\end{equation}
The packets accumulate at this bottleneck node and from there congestion
spreads over the entire network. Eq.\ (\ref{eq:mucrit}) is related to 
the end-to-end throughput via
\begin{equation}
\label{eq:Te2ebitsend}
  T_{e2e} 
    =  \mu^{\rm crit} N
    =  \mathrm{min}_i \left(
       \frac{N(N-1)}{B_i \tau_i^{\rm send}}
       \right)
       \; .
\end{equation}
This capacity-like quantity represents the maximum rate of end-to-end 
communications which can be completed successfully without network
overloading.

In Ref.\ \cite{kra04}, the blockings induced by the medium access 
control of one- and two-hop neighbors have been directly modeled into 
the sending time $\tau_i^{\rm send}$. Upon employment of a simple 
queuing behavior for the nodes' buffers, a set of coupled linear 
equations for the sending times of all nodes has been derived. Although 
this approach has let to a good qualitative understanding of the 
end-to-end throughput for various multihop ad hoc network structures, 
its analytical form is too complicated and numerically too expensive to 
be suited as an objective function. Another, much simpler description 
has to be found. 

Guidance for such a description will be given by two particular network
examples. In a fully connected network each node has direct 
communication links to all other nodes. For all nodes the degree is 
$k_i=N-1$. Initial sender and final recipient of each end-to-end 
communication are only one hop away from each other. Then, due to medium 
access control, each single one-hop end-to-end communication blocks the 
overall network, leading to a maximum throughput $T_{e2e}=1$ of one 
completed end-to-end communication per time step. Any detailed 
proposition of (\ref{eq:Te2ebitsend}) has to fulfill this $T_{e2e}=1$ for 
fully connected networks.

The second particular network example comes with a central hub, which 
has direct communication links to all other nodes. All end-to-end
communications first go from the initial sender to the central hub, and
from there to the final recipient. For simplicity, we discard the 
central hub as initial sender as well as final recipient. In each of the 
two one-hop transmissions involved in an end-to-end communication again 
the overall network is blocked by medium access control. First the 
receiving central hub blocks all its neighbors, and then all its 
neighbors are again blocked during the subsequent forwarding to the 
final recipient. This limits the maximum throughput to $T_{e2e}=0.5$, 
i.e.\ one completed end-to-end communication per two time steps. Again,
any further proposition based on (\ref{eq:Te2ebitsend}) also has to 
fulfill this $T_{e2e}=0.5$ for a central-hub network.

If all neighbors of node $i$ are currently not involved in prospected
one-hop transmissions, then $i$ is able to send its packet right away. In 
this case the sending time of (\ref{eq:muiout}) would be 
$\tau_i^{\rm send}=1$ in units of a typical one-hop packet-transmission 
time interval. For the other extreme, when all neighbors either also
attempt to forward a packet or are expecting packet receipt from some
other node, the average sending time takes on its upper limit
\begin{equation}
\label{eq:tsendki}
  \tau_i^{\rm send} 
    =  1 + k_i^{\rm in}
       \; ,
\end{equation}
where $k_i^{\rm in}$ represents the ingoing node degree. It is tempting
to insert this estimate into (\ref{eq:Te2ebitsend}), leading to 
\begin{equation}
\label{eq:Te2ebiki}
  T_{e2e} 
    =  \mathrm{min}_i 
       \left( \frac{N(N-1)}{B_i (1+k_i^{\rm in})} \right)
       \; .
\end{equation}
In fact, this expression is consistent with $T_{e2e}=1$ of a fully 
connected network, where $k_i^\mathrm{in}=B_i=N-1$. However, it is not 
consistent with $T_{e2e}=0.5$ of the central-hub network. With 
$k_{\rm c.h.}=N$ and $B_{\rm c.h.}=N(N-1)$ it is the central hub itself 
which produces the maximum of $B_i (1+k_i^{\rm in})$, leading to the 
wrong estimate $T_{e2e}=1/(N+1)$. Consequently, the ansatz 
(\ref{eq:tsendki}) for the sending time is ruled out. 

We propose another empirical ansatz for the sending time. Node $i$
will be blocked, if one node in its vicinity is involved in a packet
transmission. We have to consider all nodes that have a directed link
towards node $i$, because these block $i$ via medium access control
during their active participation in a one-hop transmission. This set 
of $i$'s ingoing neighbors is abbreviated as $\mathcal{N}_i^{in}$. As a
proposition, we set the activity of an ingoing neighbor $j$ equal to the 
relative betweenness centrality $B_j / B_i$. Then, the sending time is
estimated as
\begin{equation}
\label{eq:tsendbicum}
  \tau_i^{\rm send} 
    =  \frac{1}{B_i} 
       \left( B_i + \sum_{j\in\mathcal{N}_i^{in}} B_j \right)
       \; .
\end{equation}
It introduces the cumulative betweenness centrality
\begin{equation}
\label{eq:bicum}
  B_i^{cum} 
    =  B_i + \sum_{j\in\mathcal{N}_i^{in}} B_j
       \; .
\end{equation}
Insertion of (\ref{eq:tsendbicum}) into 
(\ref{eq:Te2ebitsend}) produces
\begin{equation}
\label{eq:Te2ebicum}
  T_{e2e} 
    =  \mathrm{min}_i \left( \frac{N(N-1)}{B_i^{\rm cum}} \right)
\end{equation}
for the end-to-end throughput.

This expression is consistent with $T_{e2e}=1$ for fully connected 
networks. In this case $B_i=N-1$ and, consequently, $B_i^{cum}=N(N-1)$. 
Moreover, expression (\ref{eq:Te2ebicum}) also agrees with the 
$T_{e2e}=0.5$ of a central-hub network. There, it is again the central
hub which yields the largest $B_i^{cum}$. Its betweenness centrality
amounts to $B_{\rm c.h.}=N(N-1)$. Each of its $N$ neighbors counts
$B_{\rm ngb(c.h.)}=N-1$. All this totals to 
$B_{\rm c.h.}^{cum} = B_{\rm c.h.} + N B_{\rm ngb(c.h.)} = 2N(N-1)$.

\begin{figure}
\begin{centering}
\epsfig{file=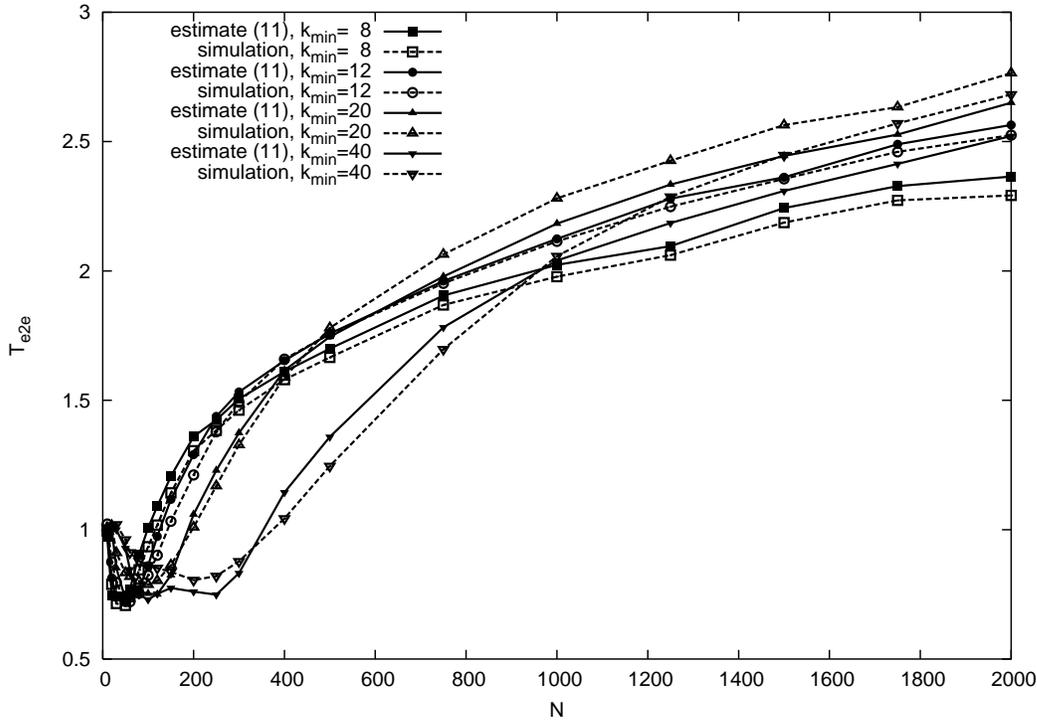,width=1.00\textwidth}
\caption{
Network-size dependent end-to-end throughput for minimum-node-degree 
networks with $k_{\rm min}=8$ (squares), $12$ (circles), $20$ 
(up-triangles), $40$ (down-triangles). Curves with full symbols 
represent the estimate (\ref{eq:Te2ebicum}). Respective curves with open
symbols result from a generic packet traffic simulation. An average over 
$100$ independent network realizations has been performed for each 
symboled $N$.
}
\label{fig:fig2}
\end{centering}
\end{figure}

Beyond these two limiting network examples, the quality of 
(\ref{eq:Te2ebicum}) is decided with the minimum-node-degree network 
structures. Generic packet traffic simulations with shortest-path
routing, as described in Ref.\ \cite{kra04}, have been used to determine 
their end-to-end throughput. For various $k_{\rm min}$ 
and in dependence of the network size $N$, these results are shown in 
Fig.\ \ref{fig:fig2}. The same network realizations, as used for the
simulations, have then been taken to determine (\ref{eq:Te2ebicum}).
The betweenness centrality based on shortest multihop paths has been 
calculated with an algorithm similar to that described in Ref.\ 
\cite{new01b}. The overall agreement between the estimate 
(\ref{eq:Te2ebicum}) and the throughput curves obtained from the generic 
packet traffic simulations is remarkable for all the various 
$k_{\rm min}$ values. This proves the high quality of expression 
(\ref{eq:Te2ebicum}), which will now serve as objective function for the 
network-structure optimization of the end-to-end throughput.

Note, that for large network sizes the results of Fig.\ \ref{fig:fig2} 
suggest a scaling law of the form $T_{e2e} \sim (N-N_0)^\gamma$. Fitted 
values for $N_0$ and $\gamma$ are given in Tab.\ \ref{tab:scaling}. 
Except for $k_\mathrm{min} > 20$, the scaling exponent 
$\gamma = \gamma(k_{\rm min})$ is found to depend only weakly on the 
minimum-node degree. For all cases it falls well inside the range 
$0 < \gamma < 0.5$. The upper estimate $\gamma < 0.5$ has been first 
given by \cite{gup00}. Despite a homogeneous end-to-end traffic pattern 
this overestimation neglects the heterogeneities in the one-hop traffic 
as a consequence of the spatial network geometry. In general, nodes in 
the spatial center of the network have to carry a higher load than nodes 
in the periphery. Taken alone, $\gamma > 0$ proves that for 
sufficiently large enough network sizes multihop networks produce a 
larger throughput than central-hub networks. However, this statement is 
much too modest, since for all curves of Fig.\ \ref{fig:fig2} and for 
all network sizes, the absolute value of the end-to-end throughput is 
always larger than $T_{e2e} = 0.5$. For network sizes slightly above 
$N_0$, the end-to-end throughput of the various minimum-node-degree 
networks becomes larger than $T_{e2e} = 1$ of a fully connected network.

\begin{table}
\begin{tabular}{l|rr|rr}
$T_{e2e} \sim (N-N_0)^\gamma$
        & \multicolumn{2}{c}{estimate (\ref{eq:Te2ebicum})}
        & \multicolumn{2}{c}{simulation} \\
                                    & $N_0$ & $\gamma$  & $N_0$ & $\gamma$\\
\hline
$k_\mathrm{min} =  8              $ & $ 24$ & $ 0.23 $  & $ 65$ & $ 0.22 $\\
$k_\mathrm{min} = 12              $ & $ 69$ & $ 0.25 $  & $127$ & $ 0.22 $\\
$k_\mathrm{min} = 20              $ & $179$ & $ 0.24 $  & $219$ & $ 0.24 $\\
$k_\mathrm{min} = 40              $ &$(400)$& $(0.22)$  &$(389)$& $(0.29)$\\
$\mathrm{opt}(k_\mathrm{min} =  8)$ & $ 36$ & $ 0.41 $  & $ 37$ & $ 0.43 $\\
$\mathrm{opt}(k_\mathrm{min} = 12)$ & $ 84$ & $ 0.40 $  & $(52)$& $(0.49)$\\
$\mathrm{opt}(k_\mathrm{min} = 20)$ & $(94)$& $(0.47)$  & $(65)$& $(0.54)$\\
$k_\mathrm{min} =   8$, $B_i^\mathrm{cum}$-metric 
                                    & $  0$ & $ 0.42 $  & $ 23$ & $ 0.41 $\\
$k_\mathrm{min} =  12$, $B_i^\mathrm{cum}$-metric 
                                    & $ 22$ & $ 0.43 $  & $ 63$ & $ 0.41 $\\
$k_\mathrm{min} =  20$, $B_i^\mathrm{cum}$-metric 
                                    & $ 93$ & $ 0.40 $  & $127$ & $ 0.41 $\\
\end{tabular}
\caption{
Scaling exponent $\gamma$ and parameter $N_0$ of the end-to-end 
throughput $T_{e2e} \sim (N-N_0)^\gamma$. Rows 1--4 are for 
minimum-node-degree networks and rows 5--7 are for the respective
structure-optimized networks, all with shortest-multihop-path routing.
Rows 8--10 are again for minimum-node-degree networks, but now with use
of the $B_i^{\rm cum}$-routing metric. Column blocks 2 and 3 represent 
the estimate (\ref{eq:Te2ebicum}) and the generic packet traffic 
simulations, respectively. In some cases, extracted parameters depend to
some extend on the interval size used for the fit; brackets indicate
these less reliable values.
}
\label{tab:scaling}
\end{table}

\subsection{Algorithmic details of optimization}
\label{subsec:twoC}

Based on (\ref{eq:Te2ebicum}), the search for optimized network 
structures is challenging. First of all, the expression for the 
end-to-end throughput depends on the network structure in a non-linear 
and non-local manner. Local addition or removal of links might change 
the end-to-end routes and thus the traffic distribution on a global 
scale. Moreover, the search space of all testable network configurations 
is very large. It is of the order $(N-1)^N$. Each of the $N$ nodes has 
its own transmission power ladder with $N-1$ rungs. Being on rung $k$ 
means that the picked node is able to reach its $k$ closest neighbors. 
Of course not all of these configurations are meaningful for wireless 
multihop ad hoc communication. Hence, it is important on the one hand to 
confine the search operations only to the meaningful ones and on the 
other hand to start with good initial network configurations. 

As initial configurations the geometric minimum-node-degree networks are 
chosen. For the moment we stick to $k_{\rm min}=8$. This sets a minimum 
node degree $k_i^{\rm min} \geq k_{\rm min}$ for each node $i$. During 
subsequent optimization operations, the respective transmission power 
values of all nodes are not decreased below their initial value, thus 
ensuring strong connectivity for all times \cite{gla03a}. Search 
operations are performed in rounds. Per round, each node is randomly 
picked once. A picked node explores in two directions. In the first move 
it increases its transmission power by one rung and, if the newly 
reached node does not already have a large enough transmission power, 
forces the latter to climb up its ladder until its rung suffices to 
successfully build a new mutual bidirectional communication link. In the 
other move the picked node steps down its transmission power ladder by 
one rung, implying that the lost neighbor might also move down its 
ladder until it reaches the rung just before another communication link 
is broken. Both moves modify the local network structure, require a 
global update of the shortest end-to-end routes and the betweenness 
centralities for all nodes, and lead to two modified estimates of the 
end-to-end throughput (\ref{eq:Te2ebicum}), which are then compared to 
the old estimate before the two explorative moves. The network structure 
yielding the largest estimate is accepted. This gradient update
procedure guarantees meaningful wireless multihop ad hoc network 
structures and keeps the occurrence of interfering one-directed links to 
a minimum. 

\begin{figure}
\begin{center}
\epsfig{file=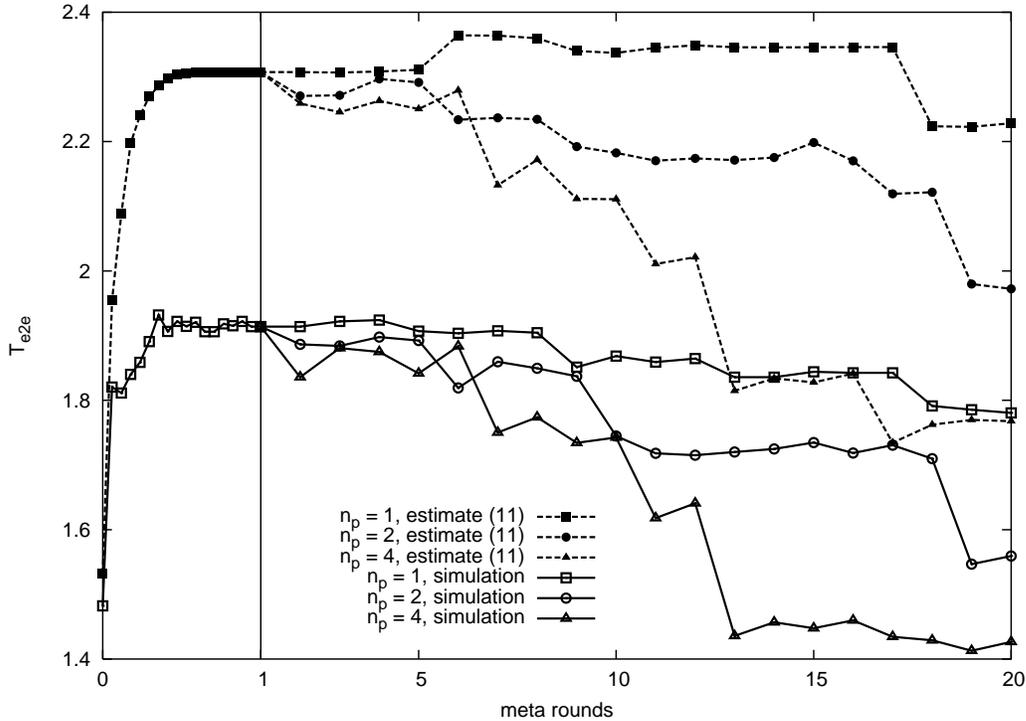,width=1.00\textwidth}
\caption{
Evolution of end-to-end throughput with progressing meta rounds. Each 
meta-round is kicked off with a random perturbation of $n_{\rm p}=1$ 
(squares), $2$ (circles), $4$ (triangles) randomly selected nodes from a 
local-maximum network configuration. Within the first meta round, the 
evolution in terms of optimization rounds is shown until the first local 
maximum of expression (\ref{eq:Te2ebicum}) is reached. A minimum-node 
degree network based on $k_{\rm min} = 8$ and $N=300$ has been chosen as 
initial network realization. The upper family of curves (with filled 
symbols) is for (\ref{eq:Te2ebicum}), whereas the lower family of 
curves (with open symbols) represents its counterpart from packet 
traffic simulations.
}
\label{fig:fig3}
\end{center}
\end{figure}

A local maximum of (\ref{eq:Te2ebicum}) is reached, if during a
complete search round no improvement of the throughput estimate is
found. Fig.\ \ref{fig:fig3} shows a typical evolution  of the 
end-to-end throughput in dependence of the number of search rounds 
until the first local maximum is reached. It only takes a modest 
number of rounds. The increase of the throughput performance is 
remarkable. Once a local maximum is reached, the respective
network realization is perturbed by forcing a small, randomly chosen 
fraction of the nodes to step up or down by one rung on their
transmission power ladder, including respective new or lost neighbor
operations as explained before. We denote the period until the next 
local maximum is found as meta-round. Fig.\ \ref{fig:fig3} also 
illustrates the evolution of the end-to-end throughput in terms of 
meta-rounds. The striking feature is that if more than one node is 
perturbed out of its local-maximum state, the throughput performance 
decreases with the number of meta-rounds. If only one node is perturbed, 
the throughput performance remains more or less the same as found for 
the first local-maximum network realization. 

If we use other minimum-node-degree networks as initial network
configurations instead of $k_{\rm min}=8$, like $k_{\rm min}=12$ and
$20$, we arrive at the same qualitative findings. Independent of
the network size, the first local throughput maximum is reached after
only a few update rounds. Subsequent meta rounds do not lead to a 
significant performance increase; on the contrary, those initiated by 
larger perturbations again result in a decrease of end-to-end 
throughput. 

It is important to check all of these results with packet traffic 
simulations. As also demonstrated in Fig.\ \ref{fig:fig3}, a strong 
correlation between the simulation results and the throughput estimate 
(\ref{eq:Te2ebicum}) is found. This is a non-trivial and important 
statement. It has been clear from the beginning that the empirical 
expression (\ref{eq:Te2ebicum}) is not fully exact. A small discrepancy
to the unknown true expression remains. A subsequent optimization with
respect to (\ref{eq:Te2ebicum}) further broadens the gap. The important 
statement is that the end-to-end throughput of the packet traffic 
simulation also increases significantly and in correlation to 
(\ref{eq:Te2ebicum}).

Taken together, the results obtained from (\ref{eq:Te2ebicum}) and from
packet traffic simulations show that the first found local maximum 
yields the largest throughput. All further maxima show a lower 
performance. This allows to terminate the optimization after the first 
meta-round. We are well aware that such an early termination most likely 
will not provide the global maximum, maybe even not a close-by network 
realization. However in view of an engineer's pragmatism, our 
maximization policy produces a well defined and fast search into a 
strong local maximum for this hard and very costly optimization problem.

\subsection{Scalability of optimized end-to-end throughput}
\label{subsec:twoD}

\begin{figure}
\begin{center}
\epsfig{file=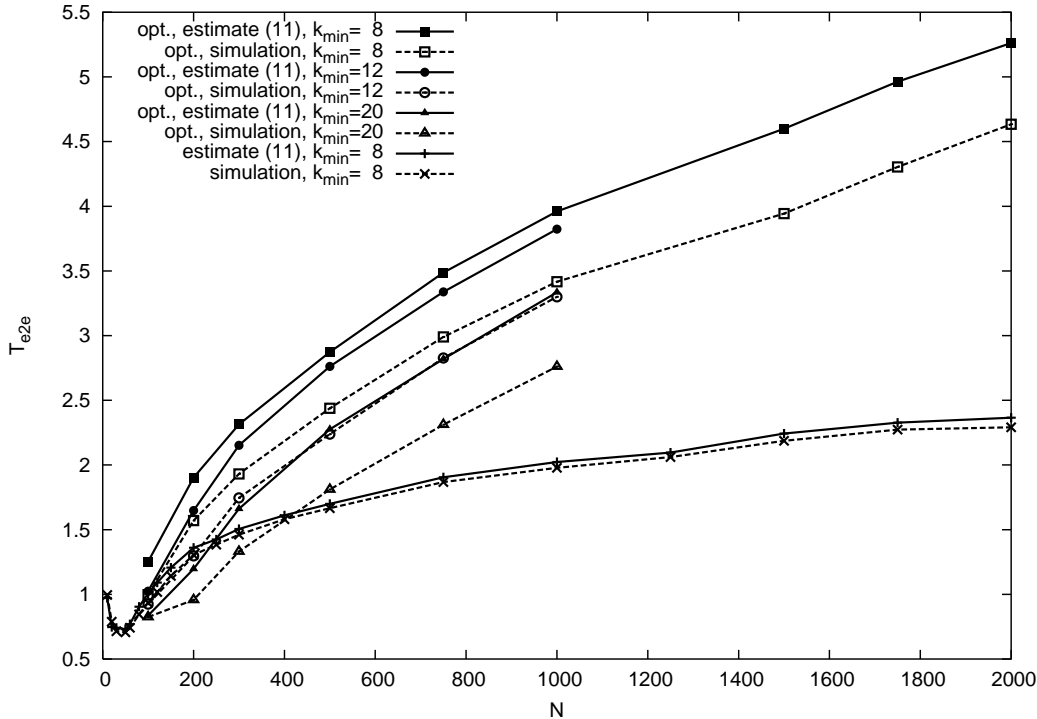,width=1.00\textwidth}
\caption{
End-to-end throughput of optimized networks as a function of network 
size, obtained from the objective function (\ref{eq:Te2ebicum}) (full 
symbols) and from generic packet traffic simulations (open symbols).
Initial minimum-node-degree networks have been chosen with 
$k_{\rm min}=8$ (squares), $12$ (circles), $20$ (triangles). An average 
over $25$ (squares, $N \leq 1000$), $9$ (squares, $N>1000$), $5$ 
(circles, triangles) independent network realizations has been performed 
for each symboled $N$. For comparison the results obtained with the 
initial $k_{\rm min}=8$ minimum-node-degree networks are again shown 
(crosses); see Fig.\ \ref{fig:fig2}.
}
\label{fig:fig4}
\end{center}
\end{figure}

For various network sizes ranging from $N=100$ to $2000$ and in 
dependence of the initial minimum-node degree $k_{\rm min}$, ensembles 
consisting of $5$ to $25$ throughput-optimized network realizations have 
been generated. Besides the optimized estimate (\ref{eq:Te2ebicum})
the end-to-end throughput has also been calculated from packet traffic
simulations. It is the numerical cost of the network structure 
optimization, which forbids larger ensemble sizes. Results are shown in 
Fig.\ \ref{fig:fig4}. 

The optimized network topologies have an end-to-end throughput 
significantly larger than their initial counterparts. The end-to-end 
throughput of the optimized topologies again reveals the scaling 
behavior $T_{e2e} \sim (N-N_0)^\gamma$ in the limit of large network 
sizes. Fitted parameter values $N_0$, $\gamma$ are given in Tab.\ 
\ref{tab:scaling}. The values found for the scaling exponent are very 
close to the upper bound $\gamma = 0.5$ given in Ref.\ \cite{gup00}.
The increase of $\gamma$ to almost $0.5$ demonstrates that within the 
optimized network topologies the heterogeneities of the one-hop traffic 
have been considerably reduced. With other words, the network structure 
has been modified in such a way that the new shortest-path end-to-end 
routes distribute the overall network traffic more evenly and reduce the 
peak traffic loads of the bottleneck nodes.

\subsection{Structural properties of optimized networks}
\label{subsec:twoE}

The optimized network structures resulting from the initial
$k_{\rm min}=8$ minimum-node-degree networks produce the largest
end-to-end throughput; consult again Fig.\ \ref{fig:fig4}. Optimized
counterparts resulting from initial $k_{\rm min}=12$ have almost the
same end-to-end throughput. However, for the larger initial 
$k_{\rm min}=20$ the respective optimized networks already come with a 
noticeably smaller end-to-end throughput. These findings are in 
accordance with the intuitive philosophy expressed in Ref.\ 
\cite{gup00}: the largest throughput is obtained once the network is 
just barely strongly connected and blockings due to medium access 
control are smallest. The minimum-node degree $k_{\rm min}=8$ just 
barely guarantees strong network connectivity and due to the small 
neighborhoods the blockings from medium access control are also small. 

\begin{figure}
\begin{center}
\epsfig{file=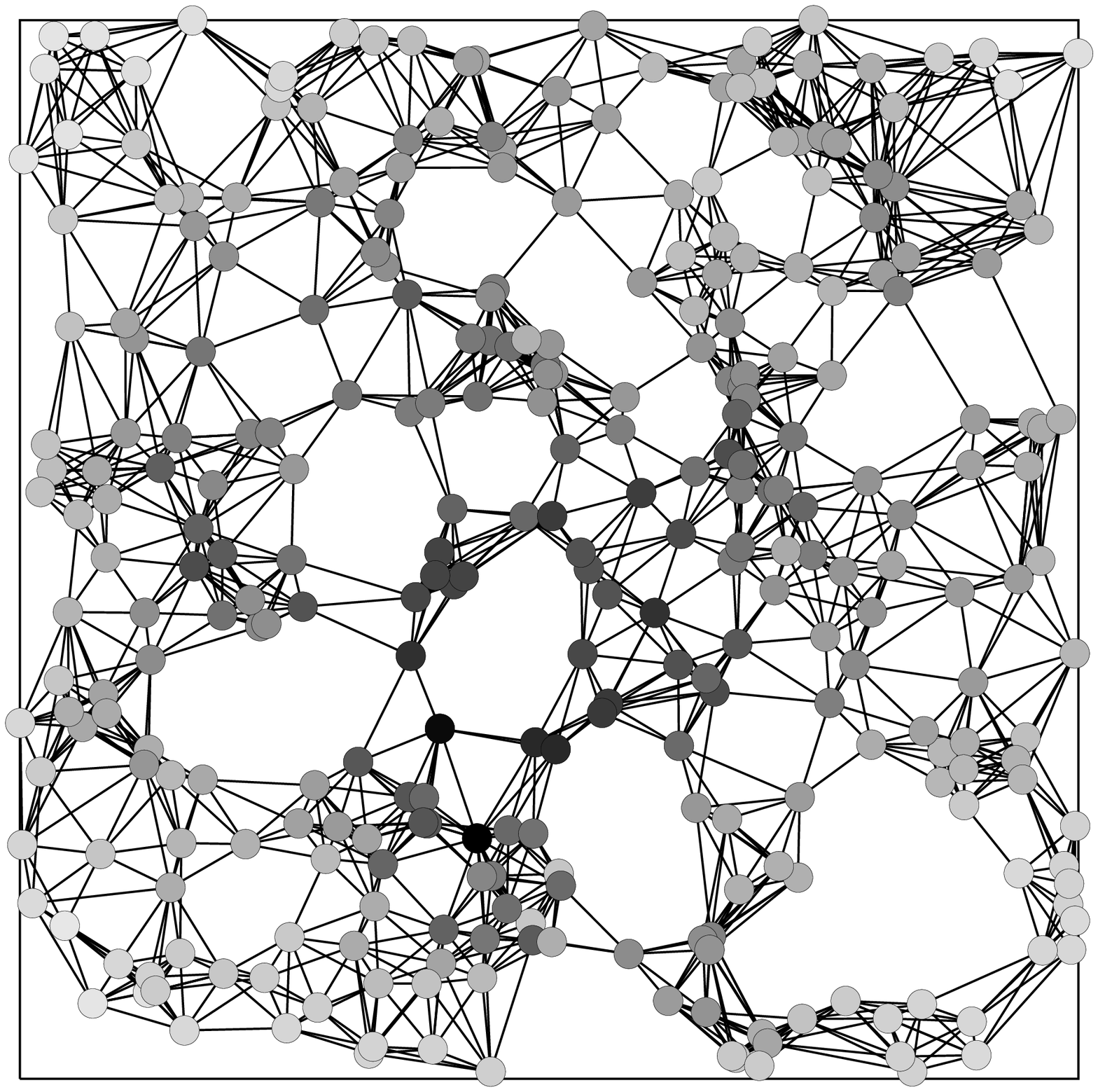,width=0.495\textwidth}
\epsfig{file=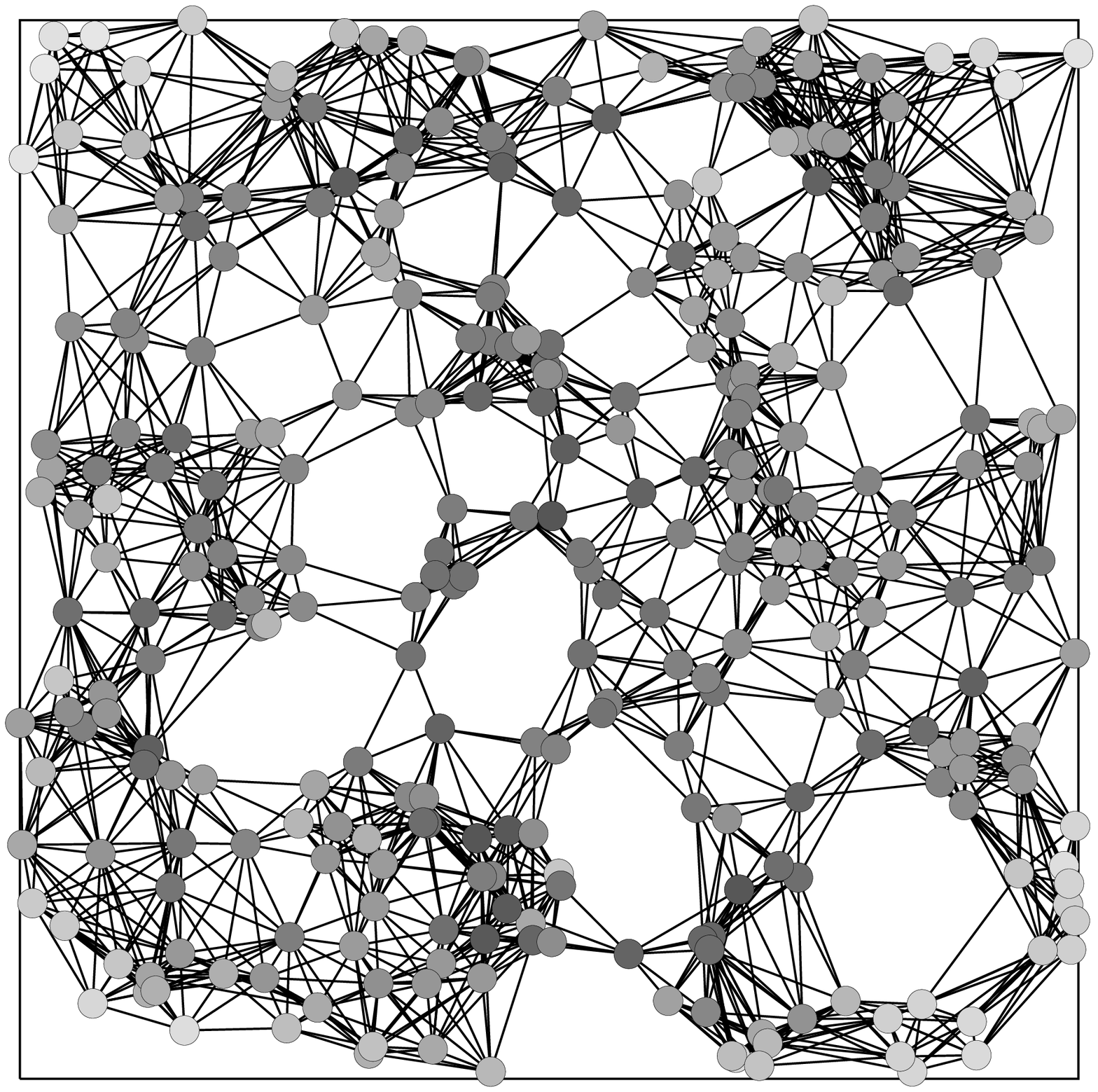,width=0.495\textwidth}
\epsfig{file=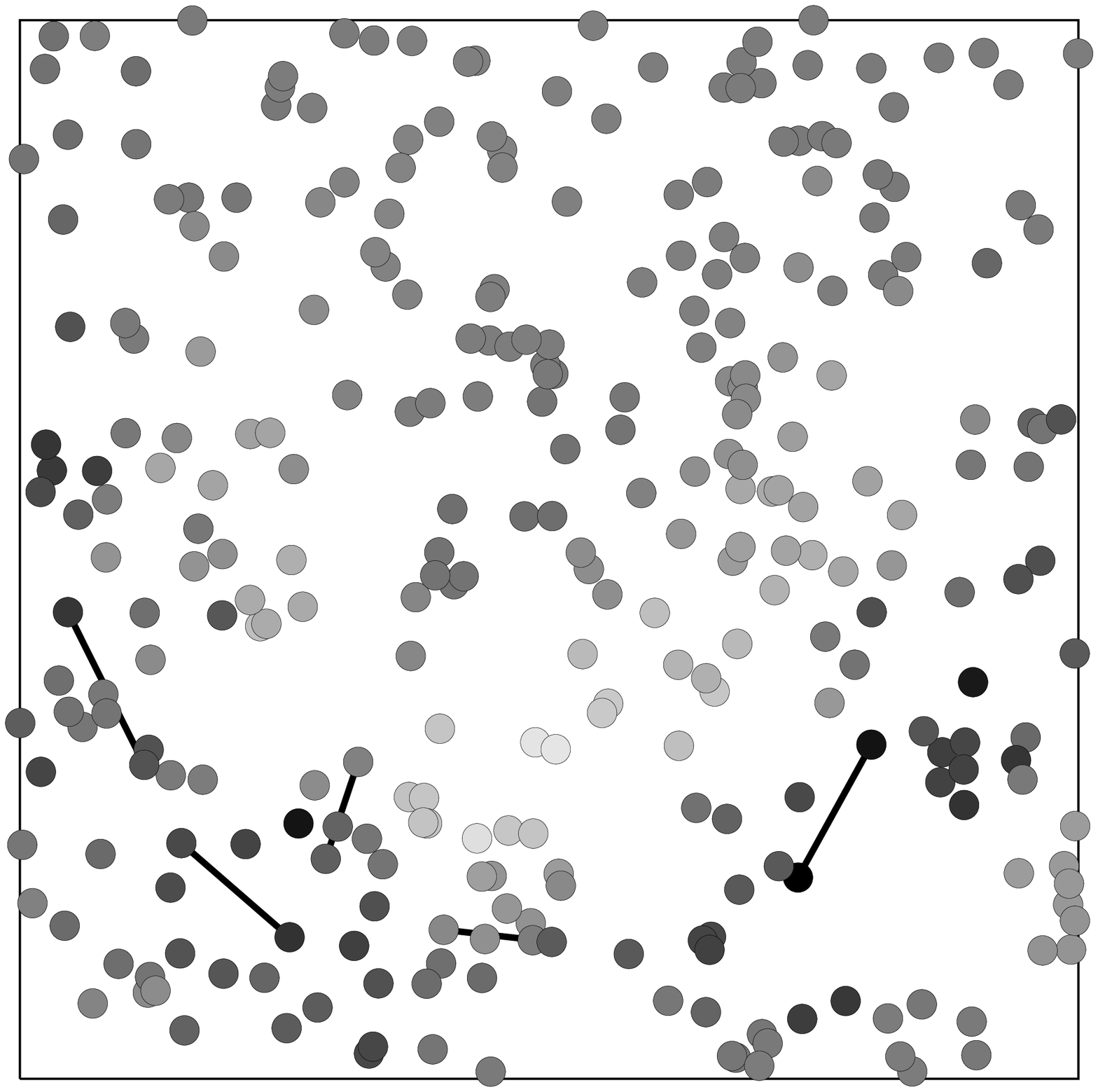,width=0.24\textwidth}
\epsfig{file=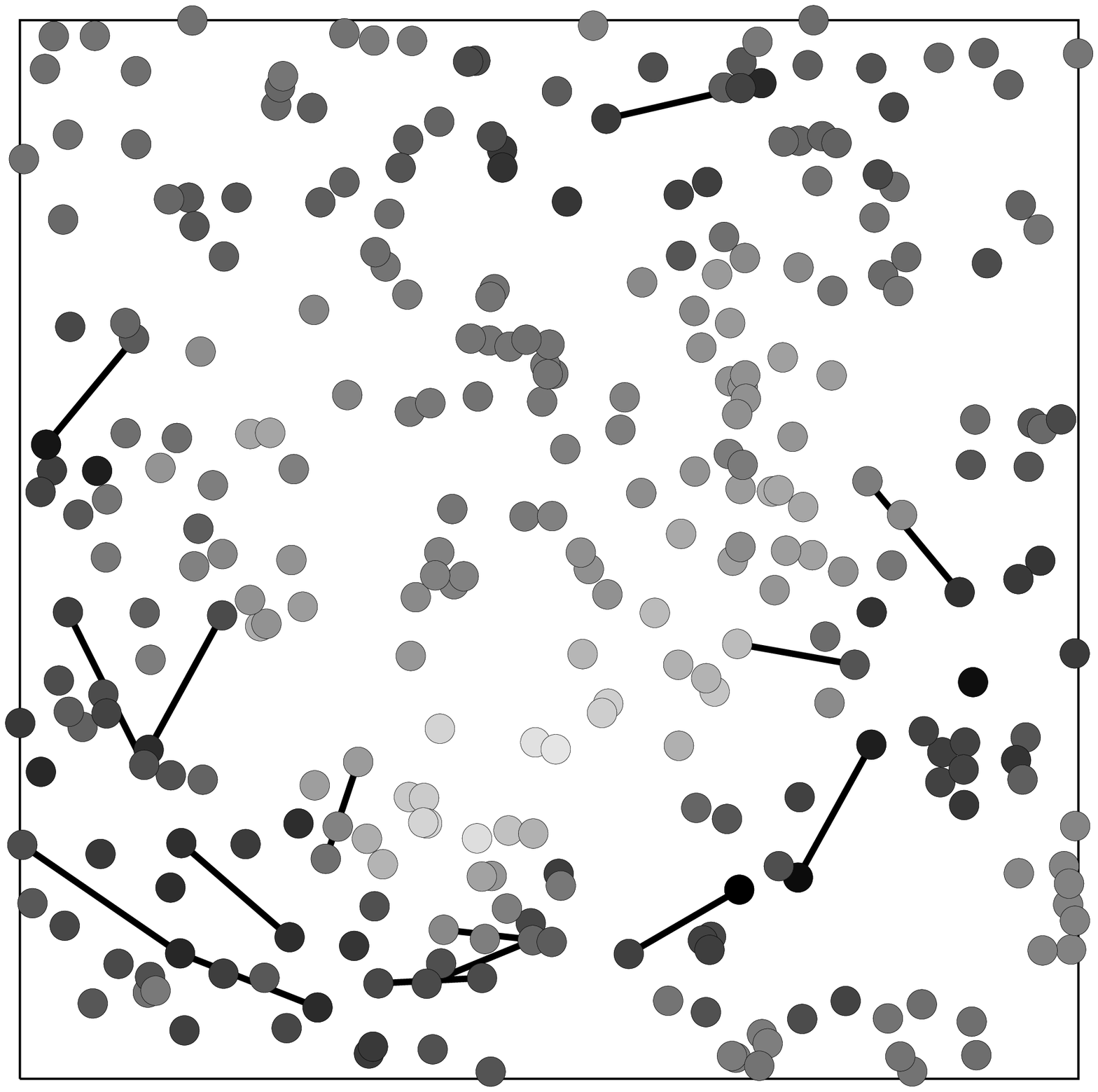,width=0.24\textwidth}
\epsfig{file=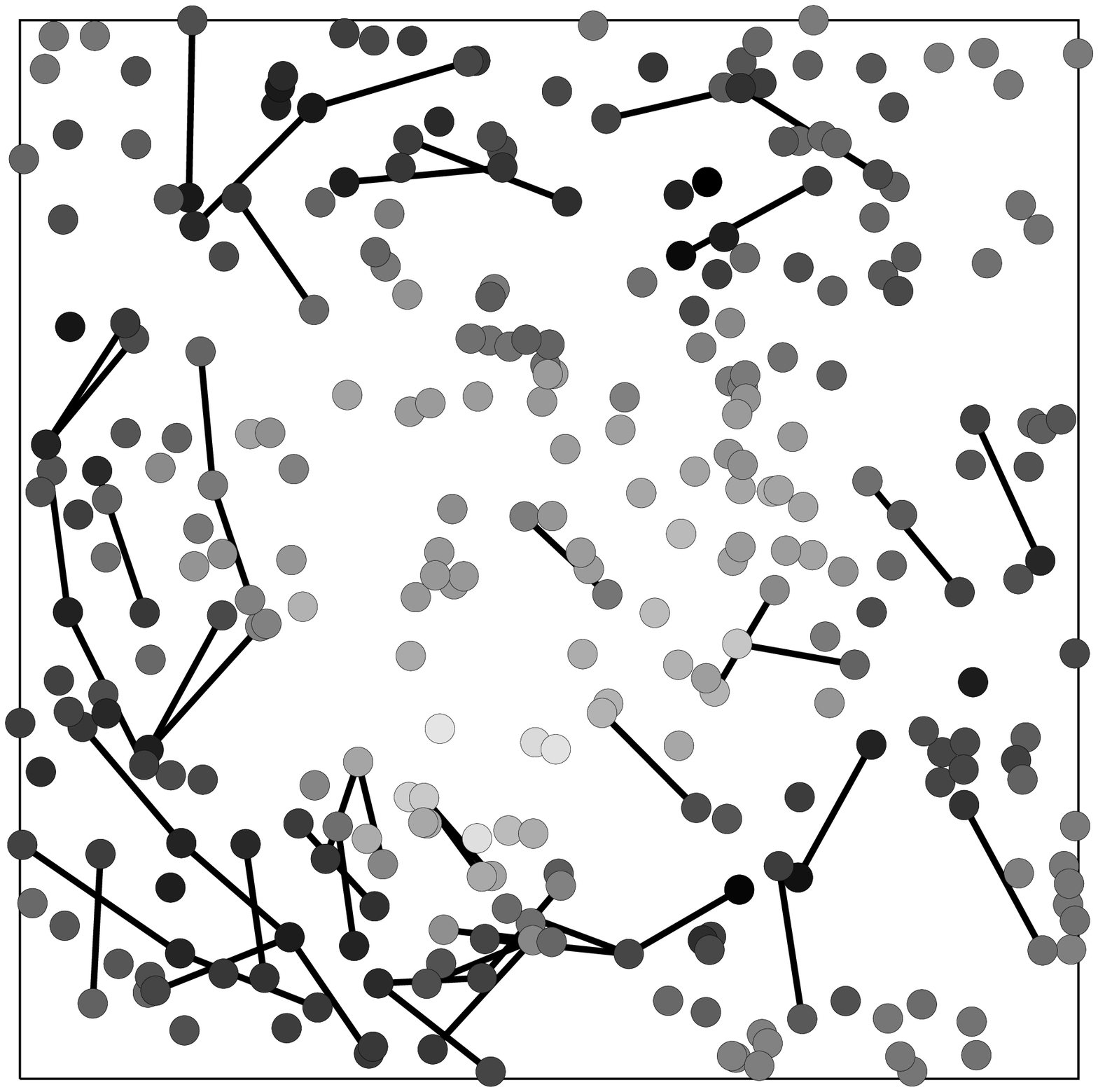,width=0.24\textwidth}
\epsfig{file=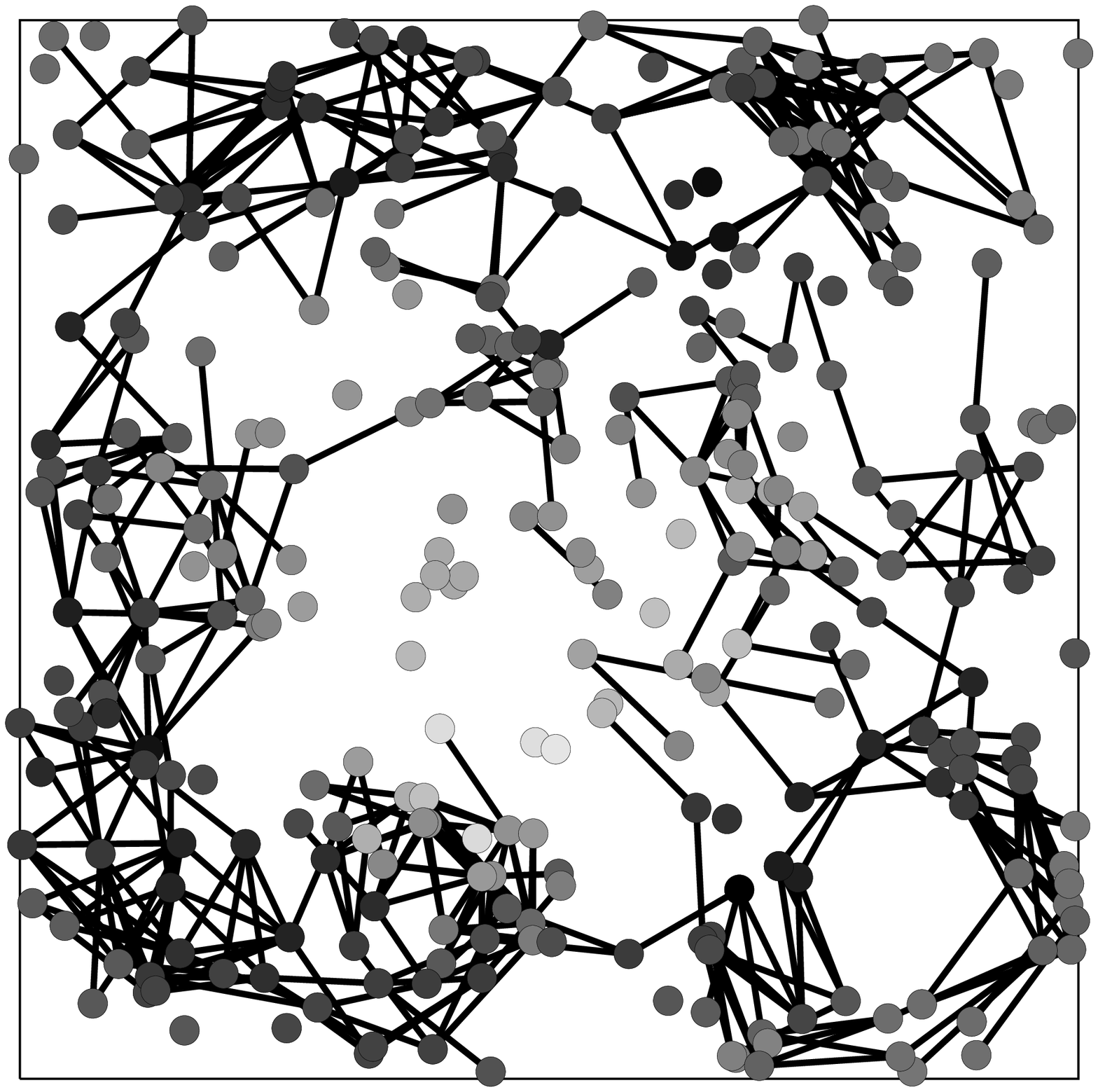,width=0.24\textwidth}
\caption{
Analysis of network structure optimization: (upper left, see also top of
Fig.\ \ref{fig:fig1}) typical initial $k_{\rm min}=8$ network 
configuration and (upper right) respective optimized network 
configuration. Only bidirectional communication links are shown; for 
reasons of readability, one-directed links have been suppressed. The gray 
scale of the nodes encodes $B_i^{cum}$, with zero (white) to maximum 
(black). From left to right, the subfigures in the lower row illustrate 
the  most important 5, 15, 50 and 391 (all) new links. Here, light 
colored nodes come with strongly reduced $B_i^{cum}$ values, whereas 
dark nodes have experienced a strong increase. 
}
\label{fig:fig5}
\end{center}
\end{figure}

Fig.\ \ref{fig:fig5}(top right) shows a typical realization of an 
optimized network. It still looks similar to the initial $k_{\rm min}=8$ 
network, which is illustrated in Fig.\ \ref{fig:fig5}(top left). For 
this example, only 391 new communication links have been added within 
the $N=300$ nodes during the optimization procedure. Fig.\ 
\ref{fig:fig5}(bottom right) shows the bidirected links added during the 
optimization. Almost none of them attaches to the spatially centered 
nodes, which are the bottleneck nodes with the largest overall 
cumulative betweenness centralities. Instead, nearly all of the new 
links are located in the greater surrounding of the most loaded nodes, 
including the outer parts of the network. The introduction of these new 
communication links modifies the shortest-path end-to-end routes in such 
a careful way that on the one hand the largest $B_i^{cum}$ values of the 
most-loaded nodes are decreased and on the other the end-to-end 
throughput performance is increased significantly. 

Not all of the 391 new links are important. A ranking of the new links 
reveals that the five most important new links are in charge of already 
49\,\% of the throughput increase, found with the generic packet traffic 
simulations. The ranking has been performed in the following way: at 
first, each of the 391 new links has been added singly to the 
minimum-node-degree network and the respective throughput 
(\ref{eq:Te2ebicum}) has been determined. The single link addition which 
yields the largest throughput defines the most important link. This link 
is then included into the network. Next, this procedure is repeated for 
the remaining 390 new links, leading to the second most important new 
link, and so on. For the 15 and 50 most important links 73\,\% and 
104\,\% of the final throughput increase are reached, respectively. For 
the remaining 341 new links the throughput increase fluctuates closely 
around 100\%. This demonstrates that a fraction of only about 50 out of 
the 391 new links is necessary to reach the optimal performance.

Similar findings hold for all other examined network realizations and 
for all network sizes. From initial $k_{\rm min}=8$ minimum-node-degree 
network to optimized network the average node degree has increased from 
more-or-less $N$-independent $\langle k \rangle = 9.9$ to $12.8$ for
$N=100$ nodes and $13.8$ for $N=2000$.

From left to right, the lower row of Fig.\ \ref{fig:fig5} shows the 
5, 15, 50 and 391 most important out of the 391 new links for this 
typical optimized network structure. The change of the nodes' 
$B_i^{cum}$ values is shown with a gray scale, where black/light gray
means increase/decrease. A close investigation of the five most 
important links reveals that two of them are formed between one- and 
two-hop neighbors of the bottleneck nodes. The other three are further 
away. All five form shortcuts in the periphery of the network. By 
reducing the hop distance between certain nodes they create new shortest 
path routes. This leads to a redistribution of traffic away from the 
highly loaded core of the network to the low-loaded periphery. For the 
15, respectively 50 most important links this behavior even becomes more 
visible. All of them are also only added in the periphery and not in the 
center of the network.

\subsection{Attempts for greedy optimization}
\label{subsec:twoF}

So far, the throughput optimization of network structure has been from a
global perspective. To become technologically more relevant, a 
distributive optimization analogue has to be found. From the previous 
Subsection we have learned that the additional, throughput-enhancing new 
links are not directly attached to the spatially centered, most-loaded 
nodes, but further away within the periphery of the network. The 
distributive control of this strong non-local dependence between network 
structure and end-to-end throughput will be facilitated for sure, once 
the nodes would have access to geographical information. However, this 
is not necessarily the case. Moreover, to some extend this also 
contradicts the advocated spirit of selforganization, since positioning 
information has to be provided to the nodes from an outside 
infrastructure. Without geographical information it will be very 
difficult for any distributive implementation of the throughput 
optimization to find the few, really important new communication links. 
We will now demonstrate our point with three different, non-geographical 
greedy-like algorithms. Of course, given the previous experience, all of 
them are based on cumulative betweenness centrality and all of them 
start with an initial $k_{\rm min}=8$ minimum-node-degree network.

\begin{figure}
\begin{center}
\epsfig{file=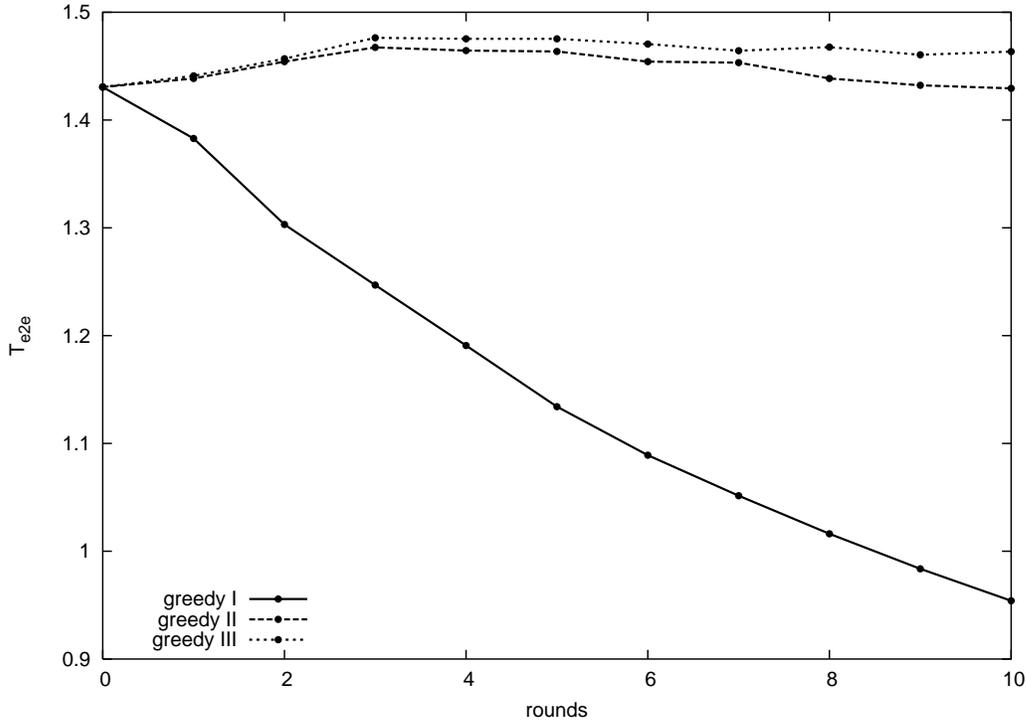,width=1.00\textwidth}
\caption{
End-to-end throughput (\ref{eq:Te2ebicum}) as a function of the number 
of greedy optimization rounds: (solid) attempt one, (dashed) attempt 
two, (dotted) attempt three; for more details, see text of Subsect.\ 
\ref{subsec:twoF}. Initially ($\mathrm{round}=0$), minimum-node-degree 
networks with $k_\mathrm{min}=8$ and $N=300$ have been used. Each 
data point is averaged over an ensemble of 100 network realizations.
}
\label{fig:fig6}
\end{center}
\end{figure}

In greedy attempt one, each single node $i$ compares its $B_i^{\rm cum}$ 
to the respective values $B_j^{\rm cum}$ of its bidirected neighbors 
$j\in\mathcal{N}_i$. Then, out of the set $\{i\}\cup\mathcal{N}_i$ it 
tags the node with the smallest value for the cumulative betweenness 
centrality. After network-wide tagging is completed, each tagged node 
increases its transmission power by one rung on the transmission power 
ladder and builds up a new bidirectional communication link, the latter
requiring that the new neighbor might be forced to step up on its own
transmission power ladder, too. This completes the first round of 
network structure change. For a couple of rounds this procedure is 
repeated. This attempt assumes that the new links, which are attached to 
the least loaded nodes, take away some of the end-to-end routes passing 
through the most loaded nodes. In this respect, on the one side the 
cumulative betweenness centrality for  the least loaded nodes would 
increase and on the other side it would decrease for the most loaded 
nodes, thus increasing the end-to-end throughput. Fig.\ \ref{fig:fig6} 
depicts the end-to-end throughput (\ref{eq:Te2ebicum}) as a function of 
the number of rounds. Right from the first round on the end-to-end 
throughput decreases.

Greedy attempt two is related to the previous one. The difference is, 
that a single node only tags itself whenever its cumulative betweenness 
is smallest compared to each of its communication neighbors. Then, again
after one round of tagging, each tagged node increases its transmission 
power to build up a new communication link, which eventually requires
also some cooperation from the new neighbor. Fig.\ \ref{fig:fig6} also
shows the respective outcome for the end-to-end throughput as a function
of the number of rounds. This time the throughput does not decrease.
However, the increase is only very modest and peaks after rounds 
$3$--$4$. 

Last in line, greedy attempt three represents yet another modification. 
In addition to attempt two, nodes are also tagged if they possess a 
local maximum in cumulative betweenness centrality, i.e.\ their 
$B_i^{\rm cum}$ is larger than that of their communication neighbors. In 
such cases these nodes then decrease their transmission power to break 
one communication link; however this is only allowed if their node 
degree does not decrease below their initial degree. Lost neighbors are 
also allowed to decrease their transmission power to the level just 
before they are about to loose another communication link. The 
motivation for this additional tagging is straightforward: if the 
most-loaded nodes loose a neighbor, then also some by-passing end-to-end 
routes might break away, eventually leading to a relief in load. Another 
quick look on Fig.\ \ref{fig:fig6} reveals that again outcome does not 
match expectations. Compared to the previous attempt, there is not much 
difference.

The outcomes of all three attempts produce a clear message. A 
distributive, non-geographical greedy-like network structure 
optimization of end-to-end throughput is a highly non-trivial and tough 
problem. Simple-minded straightforward attempts do not lead to a 
significant increase in throughput performance. Time to think about 
conceptually other ways!

\section{Performance increase with a routing metric
         based on cumulative betweenness centrality}
\label{sec:routing}

A different approach will now be taken to increase the end-to-end
throughput. In the previous Section, the optimized modification of 
network structure has been relying on a simple, fixed routing policy. 
The end-to-end communications have always been routed along the shortest 
multihop paths. What about the other way around? Keeping the network
structure fixed and modifying the end-to-end routes in such a way that
the most-loaded nodes get a substantial relief, thus increasing  
end-to-end throughput. For this endeavor we will introduce a new routing 
metric. Given the experience of the last Section, the latter will be 
based on cumulative betweenness centrality. This alternative approach 
will also allow for a distributed implementation with only moderate 
costs for coordination overhead and computation.

\subsection{Routing metric based on cumulative betweenness centrality}
\label{subsec:threeA}

In general, a routing metric is needed to determine the length of an
end-to-end route. The simplest example is the hop-count metric. In this
case the length of an end-to-end route is equal to the number of hops or 
links traversed along this route. For any pair of initial sender and 
final recipient, shortest-path routing only picks those routes which 
have the smallest multihop length. This has the disadvantage that a 
small number of nodes, especially those in the spatial center of the 
network, have to carry a large fraction of the overall network traffic, 
thus reducing the overall end-to-end throughput performance.

As a measure of a node's load we could choose its betweenness 
centrality. However, this describes the load only in terms of the 
in-packet flux; consult again Eq.\ (\ref{eq:muiin}). We prefer to take the 
product between the node's betweenness centrality and its sending time 
(\ref{eq:tsendbicum}), i.e.\ the number of end-to-end routes going 
through this node and the time it takes to send a packet. This product 
is equal to the cumulative betweenness centrality $B_i^{\rm cum}$. It is 
this cumulative betweenness centrality which we will now introduce as 
routing metric. The length of an end-to-end route between end-points 
$i$, $f$ then becomes 
\begin{equation}
\label{eq:dif}
  d_{i{\to}f}
    =  \sum_{k{\neq}f} \sigma_{i{\to}f}(k) B_k^\mathrm{cum}
       \; .
\end{equation}
All nodes $k$ belonging to the end-to-end route have 
$\sigma_{i{\to}f}(k)=1$, otherwise $\sigma_{i{\to}f}(k)=0$. The distance
(\ref{eq:dif}) sums up the betweenness centralities of all one-hop 
transmitting nodes along the end-to-end route, including the initial 
sender $i$, but excluding the final recipient $f$. The shortest 
end-to-end route between $i$ and $f$ is the one with minimum 
$d_{i{\to}f}$. 

Note, that the shortest end-to-end routes are based on the actual 
routing metric, which itself is determined by the end-to-end routes; 
consult again Eqs.\ (\ref{eq:bi}) and (\ref{eq:bicum}). Consequently, 
shortest end-to-end routes and routing metric have to be calculated in 
alternating order, until some form of convergence is reached. 

The self-consistent, iterative determination of all end-to-end routes 
subject to the $B_k^\mathrm{cum}$ metric consists of two parts:  
initialization and iteration. Initially, we set $B_k^\mathrm{cum}=1$ for 
all nodes. In one round of iterations, all nodes are picked one after 
the other. A picked node, say $i$, explores and updates all shortest 
end-to-end routes originating from itself. In doing so, it uses a 
Dijkstra-like procedure \cite{DIJ59} with the presently assigned routing 
metric. Directly after $i$'s end-to-end routing updates, the routing 
metric is also updated. All nodes in the network determine their new 
cumulative betweenness centrality. Formally, due to the decomposition 
$B_k = \sum_{i,f} B_{i{\to}f}(k)$ of the betweenness centrality, only 
the contribution $B_{i{\to}f}(k)$ resulting from all routes with initial 
sender $i$ needs to be updated. Then the next node in this round is 
picked. It already uses the freshly updated $B_k^\mathrm{cum}$ values to 
proceed further. This procedure can also be performed in a distributive 
manner. Only information about the link state of all nodes has to be 
exchanged between the nodes.

In order to fix the number of iteration rounds, generic packet traffic
simulations have been performed, as described in Ref.\ \cite{kra04}.  
Minimum-node-degree networks with $k_\mathrm{min}=8$ and $N = 30$--$300$ 
have been chosen to investigate the influence of the number of iteration 
rounds. The simulation results reveal that already two iteration rounds 
are sufficient. Beyond two rounds the end-to-end throughput does not 
increase further, although end-to-end routes are still subject to 
modifications. This defines a weak convergence. It is contrary to strong 
convergence, for which also the end-to-end routes would become stable. 
For the following we will only concentrate on the algorithm using two 
iteration rounds.

\subsection{Results on end-to-end throughput}
\label{subsec:threeB}

\begin{figure}
\begin{center}
\epsfig{file=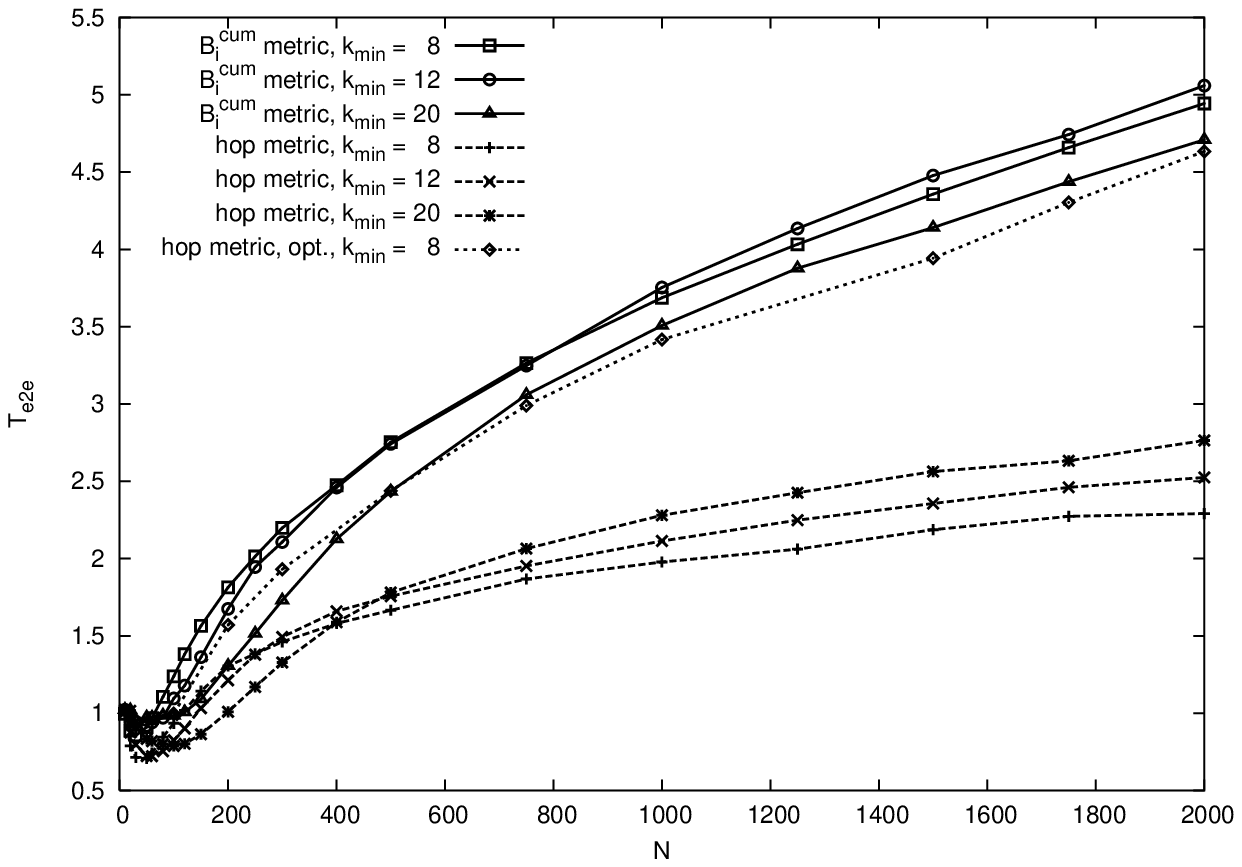,width=0.90\textwidth}
\epsfig{file=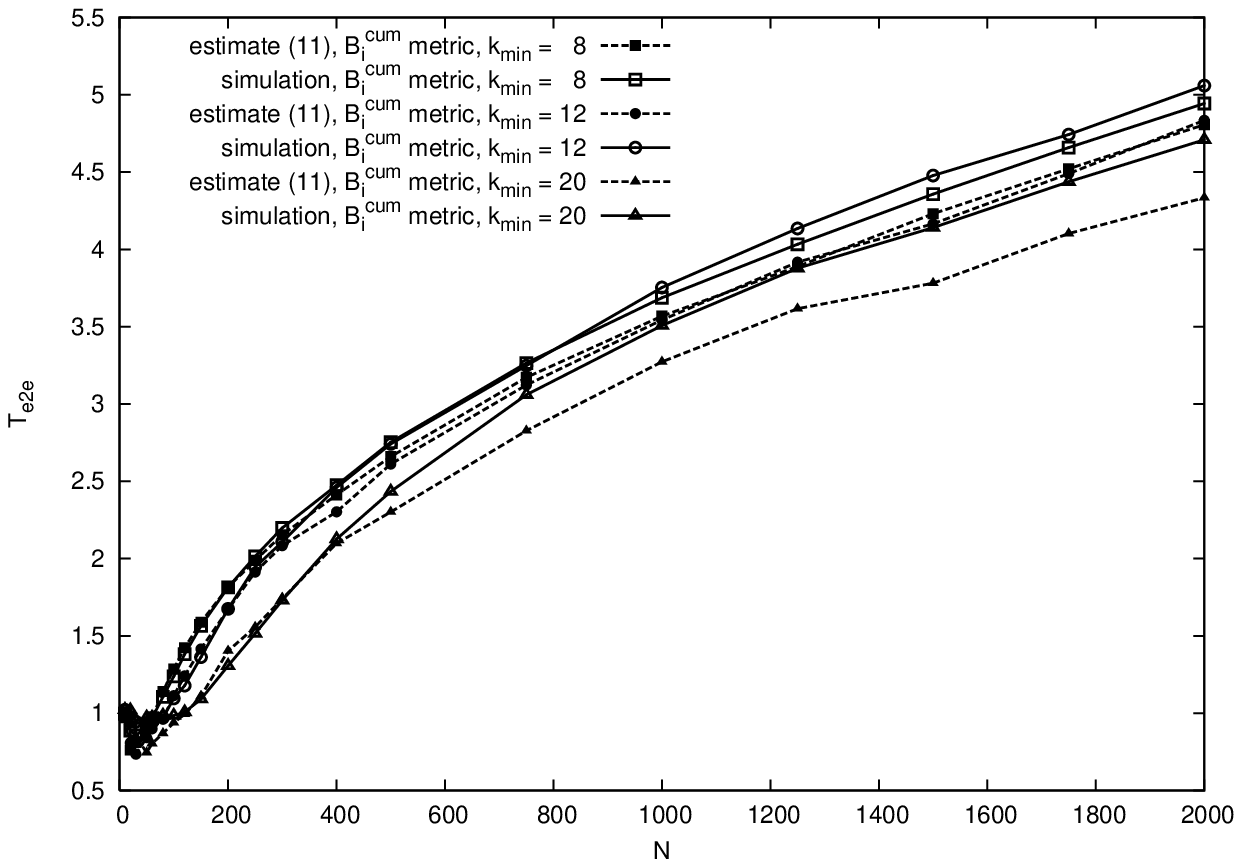,width=0.90\textwidth}
\caption{
(Top) $N$-dependent end-to-end throughput obtained with the routing 
metric based on cumulative betweenness centrality for $k_\mathrm{min}=8$ 
(open squares), $12$ (open circles), $20$ (open triangles)
minimum-node-degree networks. All curves have been determined from a 
generic packet traffic simulation. An average over $100$ (for 
$k_\mathrm{min}=8$) and $10$ (for $k_\mathrm{min}=12$, $20$) independent 
network realizations has been performed for each symboled $N$. For 
comparison, respective curves (crosses, rotated crosses, stars; see also 
Fig.\ \ref{fig:fig2}) based on the hop-count routing metric are also 
shown, as well as the analogue (open diamonds; see also Fig.\ 
\ref{fig:fig4}) resulting from the $k_\mathrm{min}=8$ network structure 
optimization. (Bottom) Comparison of the first three curves (open 
symbols) from (top) with their counterparts (closed symbols) determined 
from Eq.\ (\ref{eq:Te2ebicum}).
}
\label{fig:fig7}
\end{center}
\end{figure}

For various minimum-node-degree networks with sizes up to $N=2000$ we 
have calculated the end-to-end throughput from a generic packet traffic 
simulation, which uses the end-to-end routes obtained from the routing 
metric based on cumulative betweenness centrality. Averages over $100$
independent network realizations have been taken for $k_\mathrm{min}=8$; 
for $k_\mathrm{min}=12$, $20$ it have only been $20$. Fig.\
\ref{fig:fig7}(top) illustrates $T_{e2e}$ as a function of $N$. Again 
the scaling expression $T_{e2e} \sim (N-N_0)^\gamma$ produces a good
description for $N>200$. The resulting parameter values are listed in 
Tab.\ \ref{tab:scaling}. The scaling exponent $\gamma=0.41$ is found to 
be independent of $k_\mathrm{min}$. It is much larger than the 
respective $\gamma=0.22$-$0.24$ resulting from the hop-count metric.
Moreover, this scaling exponent is of the same order as those obtained 
from the optimized network structures based on shortest-multihop-path 
routing. In fact, on absolute scales the end-to-end throughput has 
become even slightly larger.

In principle, the expression (\ref{eq:Te2ebicum}) is not restricted to
shortest-multihop-path routing. So far, in Sect.\ \ref{sec:topology} it 
has only been tested for this case and found to produce a good estimate 
for the end-to-end throughput. A similar quality statement can now also 
be given for the routing based on the metric with the cumulative 
betweenness centrality. Fig.\ \ref{fig:fig7}(bottom) and the bottom of 
Tab.\ \ref{tab:scaling} compare the respective end-to-end throughput 
obtained from (\ref{eq:Te2ebicum}) with its counterpart obtained from 
packet traffic simulations. For the various minimum-node-degree networks
with $k_\mathrm{min}=8$, $12$, $20$ the agreement is remarkable,  
although not perfect, and of the same order as in the previous 
discussions illustrated in Figs.\ \ref{fig:fig2} and \ref{fig:fig4}. 
This proves again the high relevance of the cumulative betweenness
centrality for the generic modeling of the end-to-end communication
traffic in wireless multihop ad hoc networks. 

\begin{figure}
\begin{center}
\epsfig{file=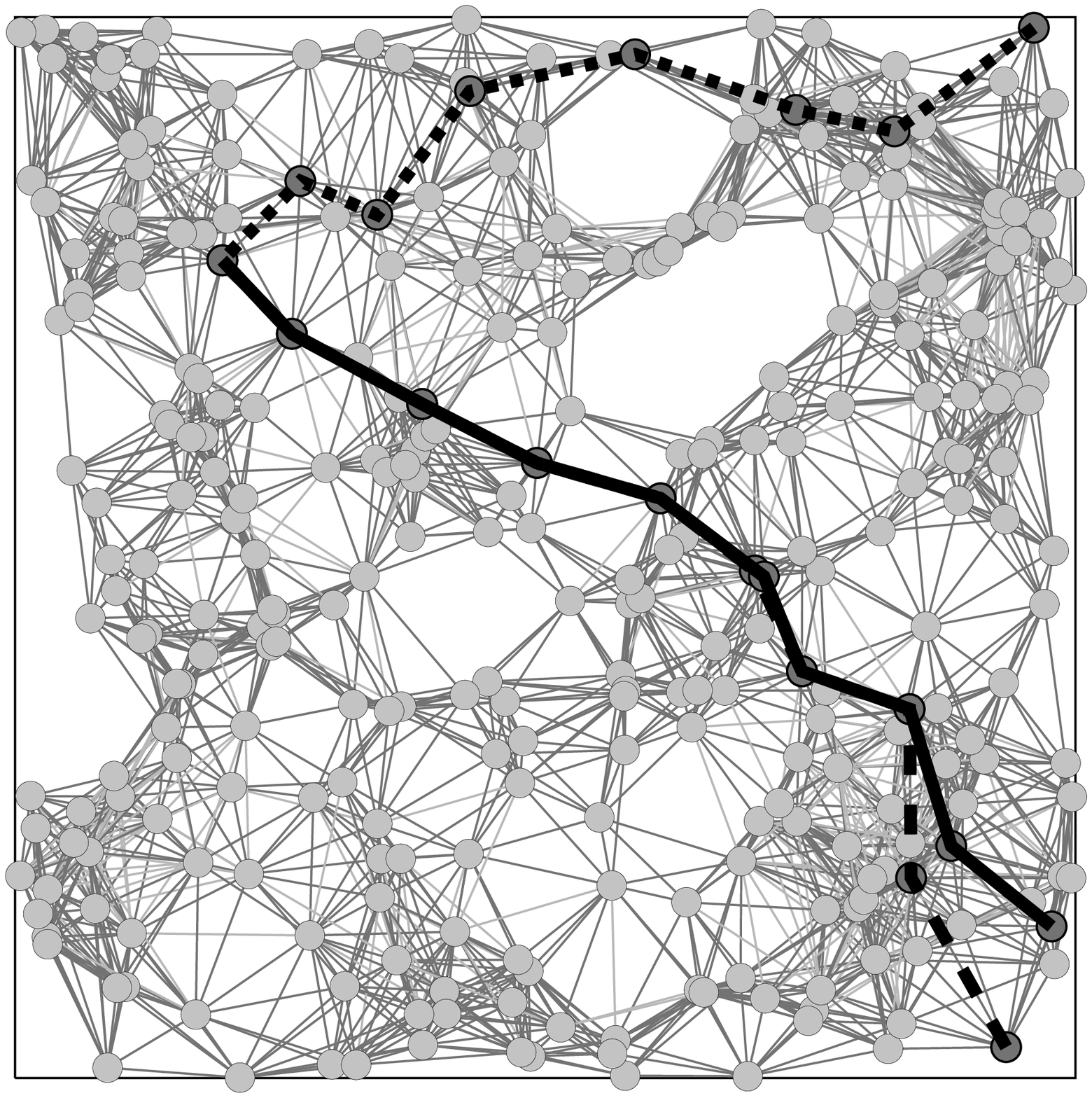,width=0.495\textwidth}
\epsfig{file=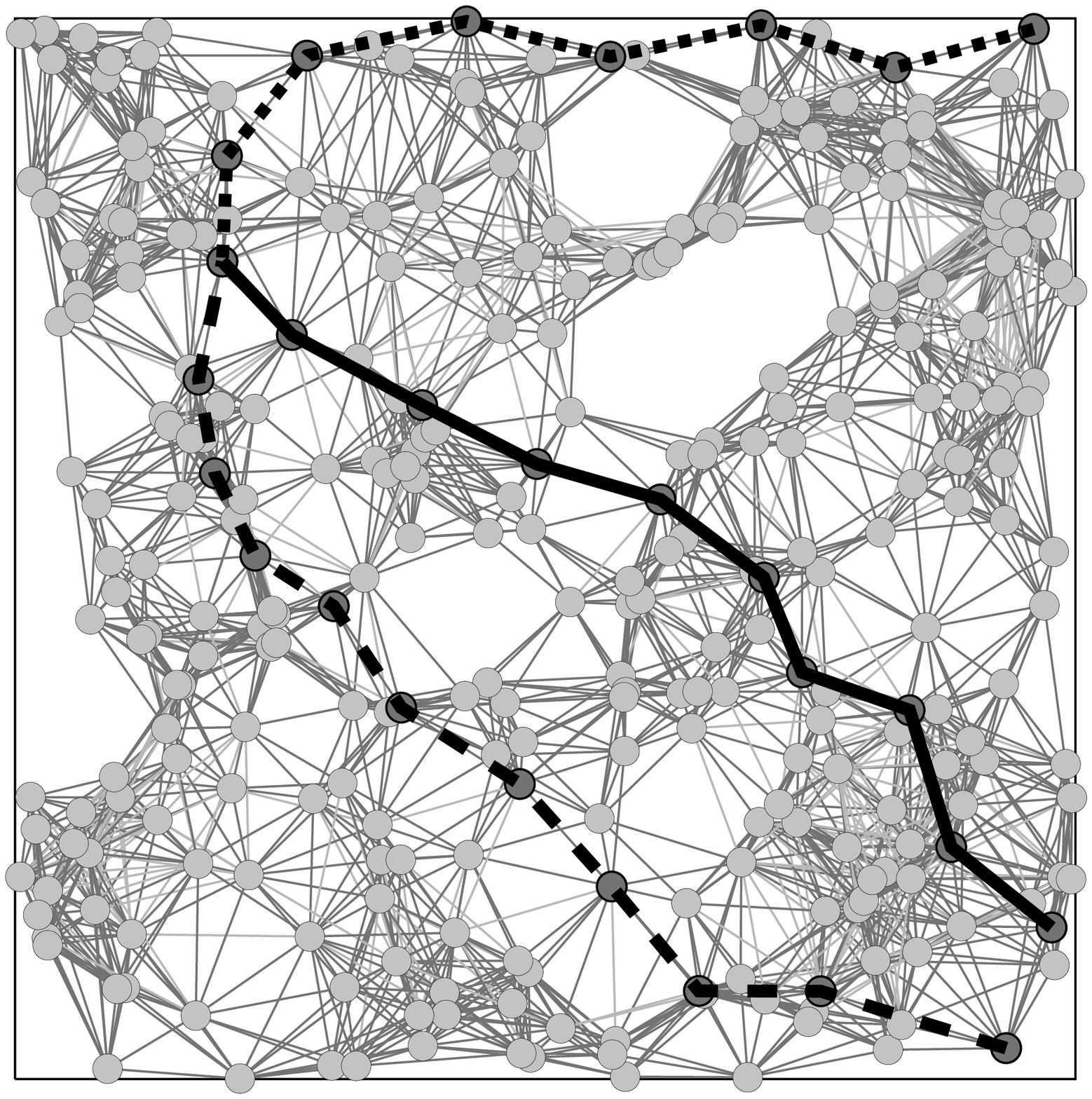,width=0.495\textwidth}
\caption{
Selected end-to-end routes based on the hop-count metric (left) and on
the metric with cumulative betweenness centrality (right). The 
underlying network is of minimum-node-degree type with 
$k_\mathrm{min}=8$ and $N=300$.
}
\label{fig:fig8}
\end{center}
\end{figure}

A consequence of the large end-to-end throughputs obtained with the 
routing metric based on cumulative betweenness centrality must be that
the end-to-end communication traffic is then more evenly distributed 
over the network. Fig.\ \ref{fig:fig8} illustrates the point. It shows 
the routes of three selected end-to-end communication partners. In case
of the hop-count metric, two of the routes are very similar. They have
the same initial node and neighboring final nodes. Both routes have a 
long common part, using the same highly congested nodes in the spatial 
center of the network. In case of the $B_k^\mathrm{cum}$ metric, these 
two routes are pushed apart. In this way they decrease the load of the 
highly congested centered nodes. Another property of the new routing 
metric is illustrated with the third end-to-end route. Whereas the 
hop-count metric chooses a geometrically rather direct path, the metric
based on cumulative betweenness centrality pushes the same end-to-end
communication to the periphery of the network. This increases the load 
of those nodes situated at the periphery to some extend. However, such
nodes are not critical to the network, and in return, the other close-by 
nodes experience a certain amount of traffic relief. 

We recapitulate: cumulative betweenness centrality describes well the 
average traffic load of nodes networked together via wireless multihop
ad hoc communication. A routing metric based on this quantity has the
strong tendency to distribute the end-to-end communication traffic 
evenly over the network. It pushes end-to-end routes towards the 
periphery. This is achieved without any geographical information and 
suffices to increase the end-to-end throughput to realms, which even 
slightly exceed those obtained with the optimization of network 
structure based on the hop count metric.

\section{Conclusion and outlook}
\label{sec:conclustion}

We have discussed wireless multihop ad hoc communication networks from 
the perspective of the Statistical Physics of complex networks. A
modification of betweenness centrality, which we have denoted as 
cumulative betweenness centrality, has been shown to be very relevant
for the intricate modeling of the end-to-end throughput performance. The
maximization of a respective objective function has led to optimized 
geometric network structures. Roughly, those are such that the networks 
are just barely strongly connected and that the size of the blockings 
resulting from the nodes' competition to gain wireless medium access are 
kept to a minimum. However, the significant increase in end-to-end
throughput depends on only a few important links. These are not attached
to the highly loaded nodes in the spatial center of the geometric 
network, but are found in the network's periphery. This pronounced 
nonlocal relationship makes it almost impossible to construct the 
optimized network structures in a technologically relevant distributive 
manner.

A second approach has been presented to increase the end-to-end 
throughput. It also relies on the cumulative betweenness centrality.
The latter is used as routing metric and leads to an iterative 
determination of end-to-end routes, which decrease the traffic load of 
the most utilized nodes. With this routing metric the end-to-end 
throughput performance of non-optimized network structures becomes even
larger than for the structure-optimized networks with the hop-count
routing metric. 

This alternative approach with the new routing metric also allows for a
technologically relevant distributed implementation. Only moderate costs
for link-state coordination overhead between the nodes and subsequent 
computation within the nodes are needed.
A performance comparison with other distributive 
implementations for routing and congestion control, like the promising 
ant algorithms \cite{caro04}, still needs to be done.

Potential further applications of the routing metric based on cumulative
betweenness centrality are not restricted to static multihop ad 
hoc and sensor networks. There is spinoff potential for more cyber 
physics related to the Internet, Ethernet and computer traffic 
engineering.

The last remark is again on network structure optimization. A very 
intriguing distributive approach could be network game theory. In Ref.\ 
\cite{ebel02} a coupling of playing games with neighboring nodes and 
network structure evolution has been introduced. Consequently, the 
optimization strategy would be: give a game to the nodes, let them play, 
and by doing so they will automatically end up in a game-dependent 
network structure, serving the optimization objective. Of course, for 
the moment this is only an idea and a lot of tough conceptual work is 
still necessary to prove it right or wrong.

\ack
W.\ K.\ gratefully acknowledges support by the Frankfurt Center for
Scientific Computing



\end{document}